\documentclass[reprint,amsmath,amssymb,aps,superscriptaddress,floatfix,prmaterials,longbibliography]{revtex4-2}

\usepackage{graphicx}
\usepackage{color}
\usepackage{booktabs}
\usepackage[caption=false]{subfig}
\usepackage{amsmath}

\newcommand*\mat{UNi$_4$B}

\begin{document}

\title{Magnetic symmetry implications of the zero- and applied-field Hall effect of \mat}

\author{Z. W. Riedel}
\affiliation{Los Alamos National Laboratory, Los Alamos, New Mexico 87545, USA}

\author{W. S. Simeth}
\affiliation{Los Alamos National Laboratory, Los Alamos, New Mexico 87545, USA}

\author{S. M. Thomas}
\affiliation{Los Alamos National Laboratory, Los Alamos, New Mexico 87545, USA}

\author{F. Ronning}
\affiliation{Los Alamos National Laboratory, Los Alamos, New Mexico 87545, USA}

\author{E. D. Bauer}
\affiliation{Los Alamos National Laboratory, Los Alamos, New Mexico 87545, USA}

\begin{abstract}
The zero-field and applied-field Hall effects in noncollinear antiferromagnets provide evidence for topological states of matter and are tied to materials' magnetic symmetry. For \mat, the antiferromagnetic state with $T_\mathrm{N}=20$~K at zero and low magnetic field is debated due to recent magnetoelectric measurements and theory work calling into question the proposed toroidal arrangement of magnetic dipole moments. For a magnetic field applied within the plane of uranium magnetic moments, the field-dependent Hall resistivity of \mat\ shows a curved response for $\rho_{yz}$ ($H{\parallel}x$, $I{\parallel}z$) and $\rho_{zx}$ ($H{\parallel}y$, $I{\parallel}x$) up to $\sim$8 T at 2 K, while an out-of-plane field results in linear behavior of $\rho_{yx}$ ($H{\parallel}z$, $I{\parallel}x$) up to 16~T. Analysis using conventional empirical relationships for the Hall effect indicate that an intrinsic effect from momentum-space Berry curvature contributes significantly to the curved transverse resistivity. Moreover, a finite zero-field Hall effect emerges at the onset of magnetic order for $\rho_{yz}$ and $\rho_{zx}$, further supporting an intrinsic origin of the Hall response. Symmetry arguments for a finite Berry curvature, an observable magnetoelectric effect, and reported magnetic structures suggest that the previously proposed magnetic space groups for the zero-field magnetic structure cannot account for the observed finite zero-field effect for two Hall orientations. Instead, we propose that $Cm'$ or $Pm'$ magnetic symmetry, depending on the parent nonmagnetic space group, is consistent with Hall resistivity, neutron diffraction, and magnetoelectric effect measurements.
\end{abstract}

\maketitle

\section{Introduction}
Noncollinear antiferromagnets can produce an anomalous Hall effect driven by intrinsic momentum-space or real-space Berry curvature \cite{nagaosa2010anomalous,suzuki2017cluster,vsmejkal2022anomalous,chen2014anomalous,ueland2012controllable,kubler2014non,surgers2016anomalous}. Intrinsic and topological effects may provide robust responses that are tunable with magnetic field, temperature, and crystallographic orientation. For example, at room temperature, the magnetic spin order of the noncollinear antiferromagnet Mn$_3$Sn produces a significant zero-field Hall response that can be manipulated with an electric current \cite{tsai2020electrical,deng2023all,xie2023efficient} and that is tunable through sample orientation with respect to an applied magnetic field \cite{nakatsuji2015large}. Similarly, Co$_{1/3}$TaS$_2$ shows a field-tunable, current-manipulable zero-field Hall effect due to scalar spin chirality \cite{park2022field,kirstein2026tunable,zhang2026current}. Topological Hall responses tied to other frustrated and/or noncollinear spin structures, such as quantum spin liquids and skyrmions, are also promising for information storage and processing \cite{machida2010time,nagaosa2013topological,liang2015current}. Ideal candidates for a tunable anomalous Hall effect will have anisotropic electronic properties, frustrated or noncollinear magnetic order, and temperature and/or magnetic field dependent magnetic order. \mat\ is a promising candidate with anisotropic behavior stemming uranium 5$f$ electrons in a pseudo-hexagonal crystal environment \cite{mentink1993reduced,mentink1994magnetic,mentink1995magnetization,mentink1997thermodynamic,bando2000large,hidaka2025magnetic}. Several subtle contributions influence the complex electronic properties and frustrated magnetic order of \mat: crystallographic distortions lowering the symmetry to orthorhombic \cite{haga2008crystal,willwater2021crystallographic,takeuchi202011b,tabata2021x}, current-induced magnetization \cite{saito2018evidence}, and a debated zero-field magnetic structure \cite{mentink1994magnetic,mentink1995antiferromagnetic,willwater2021crystallographic,ishitobi2023triple}.

The crystal structure of \mat\ was originally described with $P6/mmm$ space group symmetry and two unique uranium sites \cite{val1974new}, and later neutron diffraction experiments suggested zero-field antiferromagnetic order where two-thirds of the U magnetic moments order within the hexagonal basal plane while the other third form a paramagnetic chain along the hexagonal $c$ axis \cite{mentink1994magnetic}. The ordered moments form a toroid, a spin texture breaking time reversal and spatial inversion symmetries that can allow for a switchable ferroic moment pointing out of the uranium moment plane \cite{ederer2007towards,spaldin2008toroidal,hayami2014toroidal}. More recent X-ray and neutron crystallography studies, though, have proposed that subtle shifts of Ni and B atoms in \mat\ result in a lower orthorhombic symmetry at room temperature and below the magnetic phase transition temperature, either with $Cmcm$ space-group symmetry and four unique uranium sites \cite{haga2008crystal,takeuchi202011b,tabata2021x} or with $Pmm2$ symmetry and sixteen unique uranium sites \cite{willwater2021crystallographic}. Orthorhombic crystal field models have subsequently been used to explain a low-temperature transition near 330~mK observed in heat capacity \cite{movshovich1999second} and the elastic constant \cite{yanagisawa2021electric} that neutron Laue diffraction indicated is not a magnetic phase transition \cite{willwater2021crystallographic} and may instead be a Schottky anomaly associated with crystal-field splitting \cite{yanagisawa2021electric}.

Paired with the orthorhombic symmetry of \mat, measurements of the anisotropic current-induced magnetization \cite{saito2018evidence} led to theoretical work proposing an alternative magnetic ground state with ``triforce" magnetic dipole ordering and quadrupolar order on the remaining uranium sites \cite{ishitobi2023triple}. The triforce ground state, though, has not been definitively identified by diffraction experiments. However, other signatures may distinguish the toroidal and triforce orders. For example, Ref.~\cite{ishitobi2023triple} proposes a $Pm'm2'$ magnetic space group for the triforce order. In this case, based on symmetry rules for a finite Berry curvature, the antiferromagnetic phase may show a finite zero-field Hall effect \cite{suzuki2017cluster}. Previous studies of the Hall effect in \mat\ have focused on the zero-field nonlinear Hall effect for a single orientation, which suggests the presence of a ferrotoroidal moment in the antiferromagnetic phase \cite{oyamada2018anomalous,ota2022zero}. Here, Hall measurements are presented with three distinct Hall bar orientations to probe the symmetry of the magnetic ground state and to track the Hall effect through magnetic phase transitions observed up to 16~T.

\section{Materials and Methods}
A single crystal of \mat\ was grown using the Czochralski method in a tri-arc melt furnace. Depleted uranium, nickel (Alfa Aesar, 99.999\%), and boron (Alfa Aesar, 99.5\%) were combined in the stoichiometric 1:4:1 molar ratio for the 9~g growth charge. The charge was melted, flipped, and re-melted three times before growth. During growth, the tungsten seed rod and water-cooled Cu hearth were rotated at 20~rpm. The rod was pulled from the melt at 10~mm/h. The crystal was grown under argon purified to 8$\times$10$^{-16}$~ppm$_\mathrm{O_2}$ flowing at 1~SLM. To confirm the crystallographic orientation, the \mat\ crystal was mounted on a goniometer and oriented using a Photonic Science Laue diffractometer before direct transfer to a polisher to ensure the correct orientation. Care was taken to distinguish the hexagonal $a$ axis from the orientation 90$^\circ$ from it in the basal plane. For hexagonal symmetry, the center of the Laue pattern is the (2$\bar{1}$0) reflection when looking down $a$, the [100] direction, and is the (100) reflection when looking down $a^\ast$, the [2$\bar{1}$0] direction, making it easy to confuse the two when using the three-index notation rather than the four-index Miller-Bravais notation and possibly leading to mislabeled axes in Ref.~\cite{yanagisawa2021electric}. Single crystal X-ray diffraction, likewise, was used to confirm the Hall bar orientation after cutting and polishing.

Electrical transport data were collected with a Quantum Design Physical Property Measurement System (PPMS). Platinum leads were arranged in a six-lead geometry. The transverse (Hall) resistivity was collected with a Lakeshore resistance bridge (model 372) while the longitudinal resistivity was simultaneously collected with a Stanford Research Systems lock-in amplifier (model SR860). To eliminate the signal contribution from voltage lead misalignment, data were collected with positive and negative magnetic field after the same thermal and magnetic field history. For the Hall resistance, the signal was isolated with the formula ($R_\mathrm{H>0} - R_\mathrm{H<0}$)/2, and for the longitudinal resistance, the formula ($R_\mathrm{H>0} + R_\mathrm{H<0}$)/2 was used. 

For zero-field Hall resistivity measurements, to account for voltage lead offset and the formation of time-reversal-related magnetic domains, samples were first demagnetized above the ordering temperature with an oscillating field. Then a positive or negative 5~T field was applied perpendicular to the voltage/current plane (no overshoot on field ramp up from zero), and the sample was cooled to base temperature with the applied field. A 5~T field was selected to avoid passing through any magnetically ordered phase other than the zero-/low-field antiferromagnetic one. At the base temperature, the field was ramped down to zero, again with no overshoot to avoid a change in the magnetic field sign. Once the field settled, the transverse resistivity was measured on heating, and the Hall signal was isolated using the anti-symmetrization procedure from before [($R_\mathrm{H{\rightarrow+0}} - R_\mathrm{H{\rightarrow-0}}$)/2]. The uncertainty in the measurement was estimated from the standard deviation of four measurements at each temperature point.

Magnetization and magnetic susceptibility data were also collected with a PPMS, using the Vibrating Sample Mode (VSM) option. 
The magnetic susceptibility ($\chi=M/H$) was collected for a 0.5~T magnetic field applied along the three principal axes up to 350~K.

\begin{figure}
    \centering
    \includegraphics[width=0.9\columnwidth]{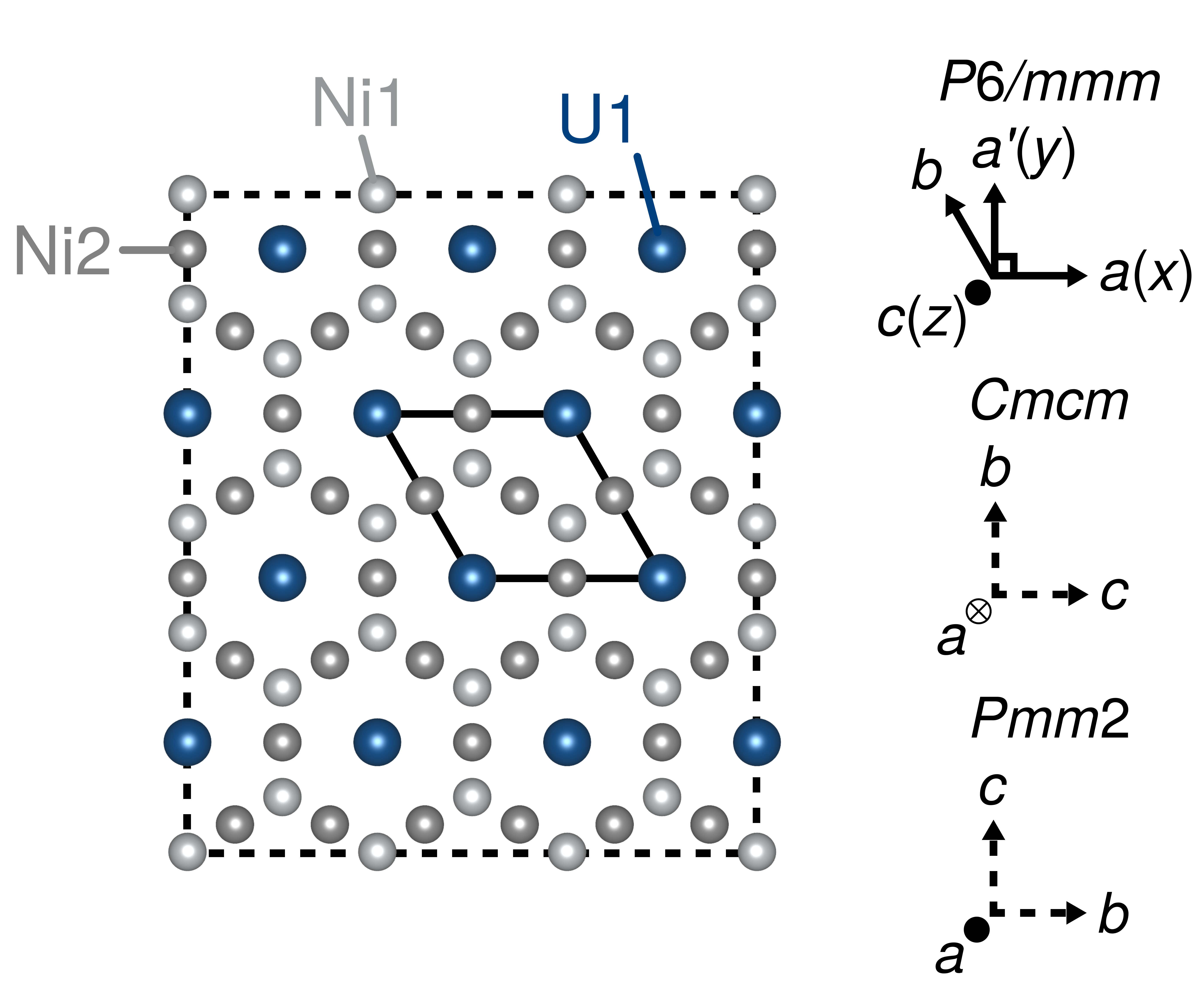}
    \caption{The unit cell of \mat\ is shown down the hexagonal $c$ axis with the undistorted hexagonal lattice positions. The orthorhombic unit cell \cite{willwater2021crystallographic} is shown with dashed lines while the hexagonal cell is shown with solid lines. Atom labels follow the hexagonal cell where Ni1 and U1 are in the same $ab$ plane, U2 sits below U1 along $c$, and boron sits below Ni1 along $c$. The $x-y-z$ axis notation used throughout the text to avoid explicit axis conversions is also indicated.}
    \label{fig:unitcell}
\end{figure}

Several notation comments deserve clarification. 
Though the crystal structure of \mat\ has orthorhombic symmetry, the hexagonal setting is frequently used because the selection of $Cmcm$ or $Pmm2$ symmetry changes the $a$, $b$, and $c$ lattice parameter definitions. 
The axes of the $P6/mmm$, $Cmcm$, and $Pmm2$ cells transform as in Fig.~\ref{fig:unitcell} with the hexagonal $a$, $a'$, and $c$ axes corresponding to the orthorhombic axes. We use $a'$ rather than $a^\ast$ because $a^\ast$, by strict definition, is dictated by the cross product $\textbf{b}{\times}\textbf{c}$, giving a direction 30$^\circ$ from $\textbf{a}$ in the hexagonal basal plane. Though six-fold symmetry dictates that the direction 90$^\circ$ from $a$ in the basal plane is equivalent, orthorhombic symmetry makes this direction inequivalent. To unify the axis definitions, we will use the $x-y-z$ notation shown in Fig.~\ref{fig:unitcell} to refer to the crystallographic unit cell, where $x=a_h=c_{oC}=b_{oP}$, $y=a'_h=b_{oC}=c_{oP}$, and $z=c_h=a_{oC}=a_{oP}$ for $h=P6/mmm$, $oC=Cmcm$, and $oP=Pmm2$. Moreover, when denoting Hall resistivity for a specific direction, we follow the subscript convention that $\textbf{E}_{i}=\boldsymbol{\rho}_{ij}\textbf{J}_{j}$, where $\textbf{E}$ is the electric field vector, $\boldsymbol{\rho}$ is the resistivity tensor, and $\textbf{J}$ is the current density vector. Therefore, $i$ is the measured-voltage direction, $j$ is the applied-current direction, and the applied-field direction is perpendicular to $i$ and $j$.

\section{Results}
\subsection{Magnetic Properties} \label{sec:mag}
\begin{figure}
    \centering
    \includegraphics[width=0.9\columnwidth]{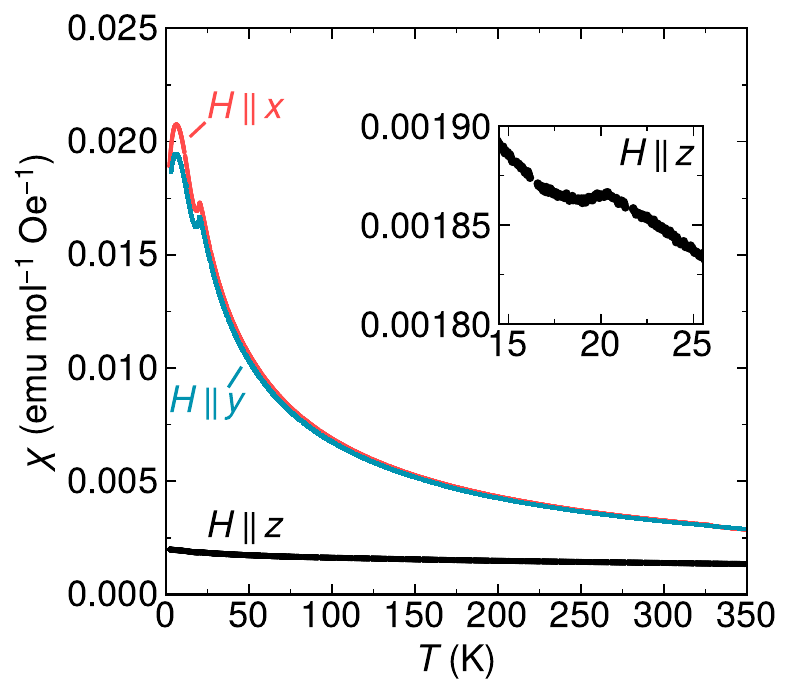}
    \caption{The magnetic susceptibility of \mat\ for a magnetic field ($\mu_\mathrm{0}H=0.5$~T) applied within the $xy$ plane ($H{\parallel}x$ and $H{\parallel}y$) and perpendicular to the $xy$ plane ($H{\parallel}z$) is shown. (Inset) The $H{\parallel}z$ susceptibility also contains an antiferromagnetic transition peaked at 20.3~K.}
    \label{fig:chi}
\end{figure}

\begin{figure*}
    \centering
    \subfloat{\includegraphics[width=0.625\columnwidth]{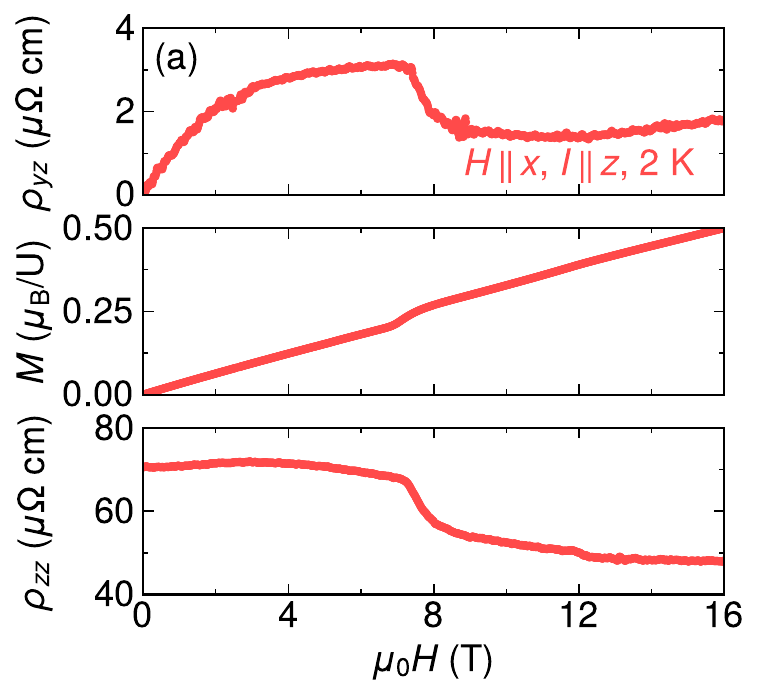}}
    \hspace{1.5em}
    \subfloat{\includegraphics[width=0.625\columnwidth]{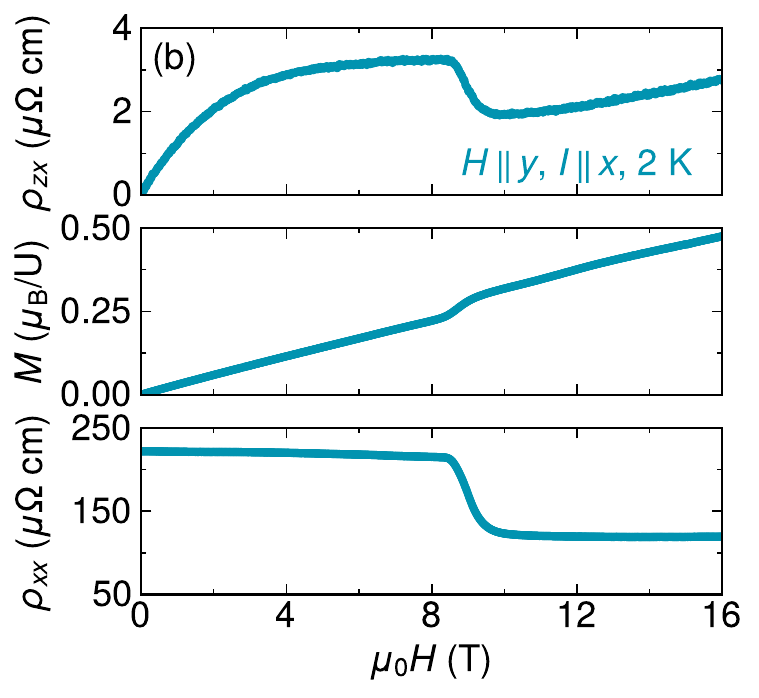}}
    \hspace{1.5em}
    \subfloat{\includegraphics[width=0.643\columnwidth]{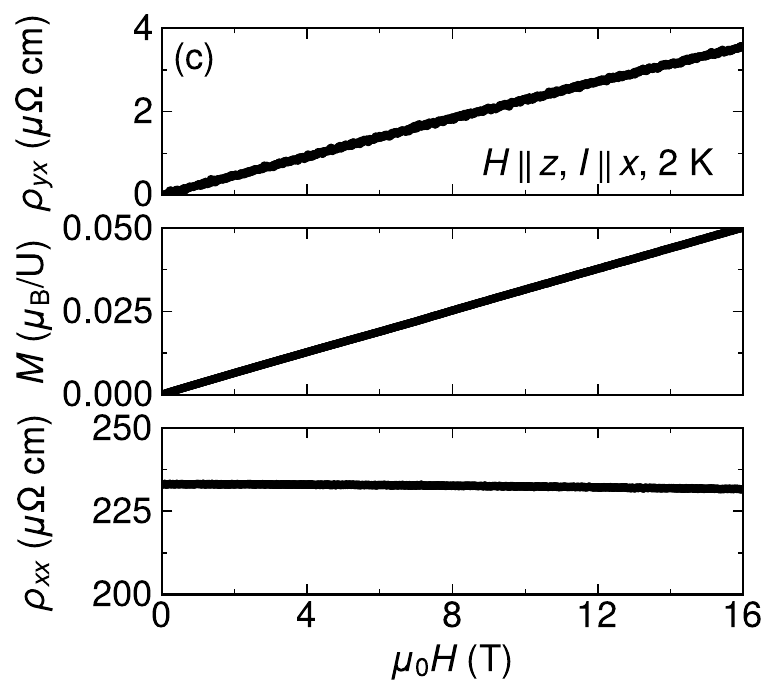}}
    \caption{The Hall resistivity (top row), magnetization (middle row), and longitudinal resistivity (bottom row) vs. $\mu_\mathrm{0}H$ of \mat\ for three magnetic field orientations are plotted for \mat. Data was collected with increasing field after zero-field cooling. The Hall resistivity for (a) $H{\parallel}x$ and (b) $H{\parallel}y$ shows significant curvature while the (c) $H{\parallel}z$ is linear in magnetic field.}
    \label{fig:panel}
\end{figure*}

The anisotropic magnetic susceptibility [$\chi(T)$] at 0.5~T (Fig.~\ref{fig:chi}) and the anisotropic magnetization [$M(H)$] at 2~K (Fig.~\ref{fig:panel}) of \mat\ match previous reports \cite{oyamada2009spin,mentink1995magnetization,mentink1993reduced,kishimoto2018magnetic,mentink1994magnetic} while showing minimal anisotropy between the $\chi(T)$ data for $H{\parallel}x$ and $H{\parallel}y$. Additional $M(H)$ data at 2, 5, 10, 15, and 25~K is provided in the Supplemental Material \cite{supplemental}.
The downturn in $\chi(T)$ at $T_\mathrm{N}=20.3$~K corresponds to the onset of long-range antiferromagnetic (AFM) order and even appears with $H{\parallel}z$, which corroborates a recent neutron diffraction study \cite{schroder2024magnetic} and a possible bump in previous sparser data \cite{mentink1994magnetic}. For all three directions, a broad hump peaked at 6.1~K is observed below $T_\mathrm{N}$, which decreases in magnitude with increasing applied magnetic field \cite{mentink1994magnetic,mihalik1996crystal}. Note that the axis scale of Fig.~\ref{fig:chi} prevents the observation of the hump with $H{\parallel}z$.

For the zero-field cooled $M(H)$ at 2~K, the expected transitions are observed. With $H{\parallel}x$, a prominent transition appears at 7.1~T (defined at the peak in $dM/dH$) while the second transition at 11.7~T is more subtle than in the previous report \cite{mentink1995magnetization}. For $H{\parallel}y$, a single transition is observed at 8.7~T. 
Not observed in the previous high-field data, which had significantly fewer data points, is a hysteretic transition for $H{\parallel}z$ (Fig.~S1 inset \cite{supplemental}). The transition is not clear after zero-field cooling, but appears at 7.3~T on decreasing field from 16~T and at 7.8~T on increasing field from -16~T and, therefore, may stem from domain physics. A nearly imperceptible shift also occurs in the Hall resistivity discussed below. On returning to zero field, a remnant magnetization is present for $H{\parallel}y$ and $H{\parallel}x$ but not for $H{\parallel}z$ [Fig.~\ref{fig:zerofield}(b)]. No remnant magnetization is present above $T_\mathrm{N}$.

\subsection{Resistivity}
\begin{figure}
    \centering
    \includegraphics[width=0.85\columnwidth]{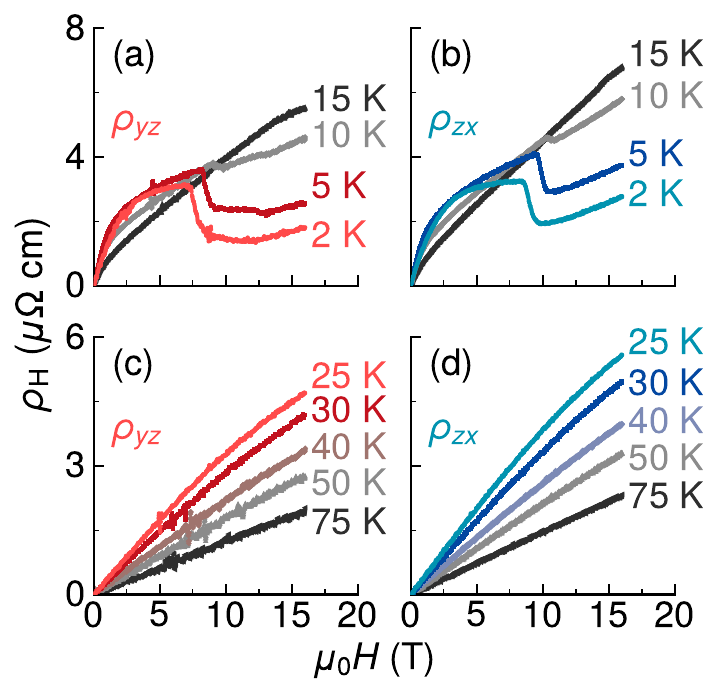}
    \caption{The field-dependent Hall resistivity in various configurations with a magnetic field applied within the $xy$ plane [$\rho_{yz}(H{\parallel}x)$, $\rho_{zx}(H{\parallel}y)$] shows (a,b) a reduction of the Hall resistivity curvature up to 20~K in the low-field AFM phase with increasing temperature. (c,d) Above 20~K, a slight curvature at high fields indicates multiband behavior that is suppressed with increasing temperature.}
    \label{fig:rhoH_multiT}
\end{figure}

Simultaneous measurement of the field-dependent longitudinal and transverse resistivity provided six datasets:
$\rho_{yz}$ and $\rho_{zz}$ with $H{\parallel}x$; $\rho_{zx}$ and $\rho_{xx}$ with $H{\parallel}y$; and $\rho_{yx}$ and $\rho_{xx}$ with $H{\parallel}z$. The zero-field cooled 2~K data are presented in Fig.~\ref{fig:panel} with additional longitudinal data at higher temperatures in the Supplemental Material \cite{supplemental}. Two additional orientations were collected at 2~K for accurate conversion from resistivity to conductivity with field applied in the $xy$ plane: $\rho_{yy}$ with $H{\parallel}x$ and $\rho_{zz}$ with $H{\parallel}y$ (Fig.~S5 \cite{supplemental}). 
Where overlapping, the longitudinal magnetoresistance data are consistent with Refs.~\cite{mentink1993reduced,mentink1995magnetization}. The center of the AFM-phase magnetoresistance bump for $H{\parallel}x$ and $H{\parallel}y$ [Fig.~\ref{fig:panel}(a)] corresponds to an observed elastic constant anomaly \cite{yanagisawa2021electric} and was initially proposed to relate to fluctuations in the one-dimensional chains of paramagnetic uranium sites \cite{mentink1995magnetization} but may instead be associated with the onset of quadrupolar order \cite{yanagisawa2021electric}. When \mat\ transitions to the first spin-reorientation phase for $H{\parallel}x$ and $H{\parallel}y$, the longitudinal resistivity undergoes a large, hysteretic drop, indicating that some uranium moments align, decreasing magnetic scattering. For the second transition observed for $H{\parallel}x$, the drop is less pronounced, indicating a subtle shift in the uranium moments, consistent with minimal change in $M(H)$.
The field-dependent Hall resistivity has a prominent curvature at 2~K in the AFM phase [Fig.~\ref{fig:panel}(a,b)] that flattens with increasing temperature [Fig.~\ref{fig:rhoH_multiT}(a,b)]. In contrast, $\rho_{yx}(H{\parallel}z$, 2~K) is linear outside of a small shift $\sim$7--8~T [Fig.~\ref{fig:panel}(c)]. For temperatures just above $T_\mathrm{N}$, $\rho_\mathrm{H}$ has a slight curvature over the range [0,16]~T but becomes linear as the temperature increases [Fig.~\ref{fig:rhoH_multiT}(c,d)].

\begin{figure}
    \centering
    \includegraphics[width=0.85\columnwidth]{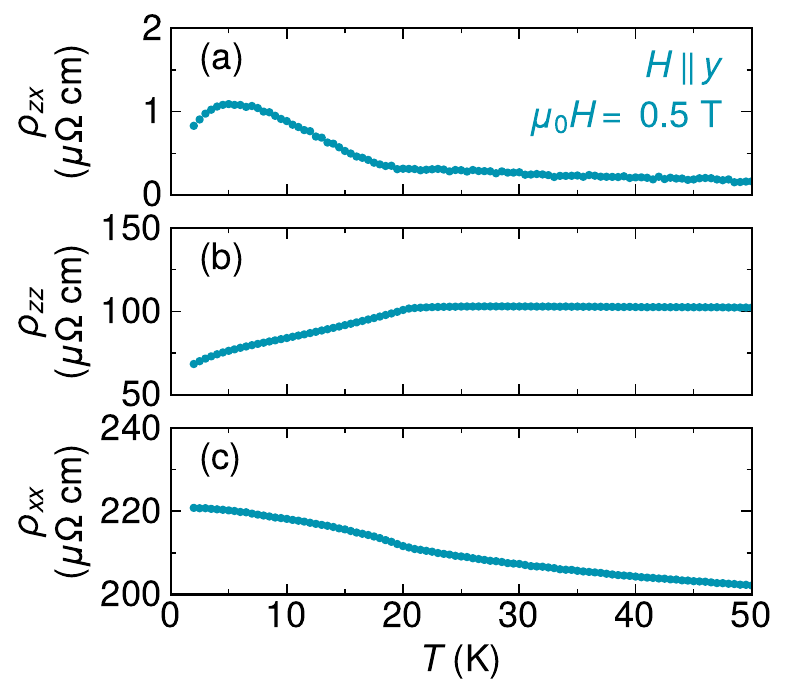}
    \caption{The temperature dependence of the Hall and longitudinal resistivity is presented for (a) $\rho_{zx}$, (b) $\rho_{zz}$, and (c) $\rho_{xx}$ with $H{\parallel}y$.}
    \label{fig:rho_ca_cc_aa_T}
\end{figure}

Figure~\ref{fig:rho_ca_cc_aa_T} shows the temperature dependence of $\rho_{zx}$, $\rho_{zz}$, and $\rho_{xx}$ ($H{\parallel}y$, $\mu_\mathrm{0}H=0.5$~T). While $\rho_{xx}$ increases with decreasing temperature, $\rho_{zz}$ decreases (c.f. Refs.~\cite{mentink1994magnetic,ota2022zero}), likely from anisotropic scattering due to the frustrated in-plane AFM order. $\rho_{zx}$($T$) shows a peak near 5~K, consistent with $\rho_{zx}$($H$), where the low-field $\rho_{zx}$ is larger at 5~K than 2~K [Fig.~\ref{fig:rhoH_multiT}(b)]. At zero applied magnetic field, a finite, temperature-dependent Hall signal appears below $T_\mathrm{N}$ for $\rho_{yz}$ and $\rho_{zx}$ but not for $\rho_{yx}$ [Fig.~\ref{fig:zerofield}(a)]. For $\rho_{yz}(H=0)$, the resistivity shifts near 10~K before further increasing. The shift matches the elastic constant change near zero-field for $H{\parallel}x$ \cite{yanagisawa2021electric}. 

\section{Discussion}
\subsection{Possible Hall resistivity contributions} 
Empirically, the Hall resistivity may be modeled with two primary contributions: the ordinary Hall contribution due to the magnetic field dependence and the anomalous Hall contribution ($\rho^\mathrm{AH}$) as in Eq.~(\ref{eq:Hall}).
\begin{equation}
    \label{eq:Hall}
    \rho_\mathrm{H} = R_\mathrm{0}\mu_\mathrm{0}H + \rho^\mathrm{AH}
\end{equation}
The curvature of the AFM-phase Hall resistivity for two orientations is not present in the higher-field phases, indicating a nonordinary contribution from the AFM order.
The form of $\rho^\mathrm{AH}$ may consist of terms proportional to the magnetization and skew scattering (${\propto}\mu_\mathrm{0}M\rho_{ii}$ where $\rho_{ii}\equiv$ longitudinal resistivity), magnetization and side-jump scattering (${\propto}\mu_\mathrm{0}M\rho_{ii}^2$), momentum-space Berry curvature, or scalar spin chirality ($\textbf{S}_i {\cdot} \textbf{S}_j{\times}\textbf{S}_k$) \cite{nagaosa2010anomalous}. A Berry-curvature-driven intrinsic contribution, like side-jump scattering, is expected to have a $\rho_{ii}^2$ dependence \cite{nagaosa2010anomalous}. Skew scattering and scalar spin chirality are not expected to dominate in \mat. Skew scattering generally dominates in the high-purity, high-conductivity regime ($\sigma_{ii}>10^6$~$\Omega^{-1}$~cm$^{-1}$) \cite{nagaosa2010anomalous}. However, $\sigma_{yy}$($H{\parallel}x$), $\sigma_{xx}$($H{\parallel}y$), $\sigma_{zz}$($H{\parallel}x$), and $\sigma_{zz}$($H{\parallel}y$)
 are all of order 10$^3$-10$^4$~$\Omega^{-1}$~cm$^{-1}$ (Fig.~S6 \cite{supplemental}), placing \mat\ in the regime where intrinsic contributions frequently dominate. Further, in the dominant skew-scattering regime, the Hall conductivity, $\sigma_{ij}$, scales linearly with the longitudinal conductivity, $\sigma_{ii}$. When the temperature-dependent resistivity is converted to conductivity following Eqs.~(\ref{eq:sig-H}) and (\ref{eq:sig-L}) [$i$, $j$, and $k$ are mutually orthogonal], $\sigma_{xz}$ does not scale linearly with $\sigma_{xx}$ in the AFM region (Fig.~S6 \cite{supplemental}). 

 \begin{equation}
    \label{eq:sig-H}
    \sigma_{ij}(H{\parallel}{k})=\frac{\rho_{ji}}{\rho_{jj}\rho_{ii}+(\rho_{ji})^2}
\end{equation}

\begin{equation}
    \label{eq:sig-L}
    \sigma_{ii}(H{\parallel}{k})=\frac{\rho_{jj}}{\rho_{jj}\rho_{ii}+(\rho_{ji})^2}
\end{equation} 
 
The second negligible contribution is scalar spin chirality. The magnetic moments of \mat\ under zero applied magnetic field are considered coplanar in previous neutron diffraction studies \cite{mentink1994magnetic,willwater2021crystallographic}, and the high-field phases for an applied field in the $xy$ plane were also predicted to be coplanar using an $XY$ mean-field model and magnetization data \cite{mentink1995magnetization}. Therefore, we assume that significant scalar spin chirality will be generated only if a magnetic field is applied out of the $xy$ plane and leads to significant canting of the U magnetic moments toward the $z$ axis. However, even when a $z$-axis field is applied, which should promote such canting, no significant curvature in $\rho_{yx}$ is observed up to 16~T (Fig.~\ref{fig:panel}c), indicating any scalar spin chirality contribution is insignificant. We are, therefore, left with a subset of plausible origins for the curved AFM phase $\rho_\mathrm{H}$: side-jump scattering, an intrinsic contribution, or multiband behavior that invalidates the simple $R_\mathrm{0}\mu_\mathrm{0}H$ estimate for the ordinary Hall effect.

Above $T_\mathrm{N}$, $\rho_\mathrm{H}$ is positive, suggesting dominant hole carriers, consistent with previous thermopower measurements \cite{mentink1995magnetic,bando2000large}. Single-band fits to data above 50~K, where the data are linear up to 16~T, are shown in Fig.~S9 \cite{supplemental}, giving temperature-dependent $R_\mathrm{0}$ values and carrier concentrations. Between 20 and 50~K, a single-band approximation no longer applies above $\sim$6-10~T where $\rho_\mathrm{H}$ deviates from a linear relationship [Fig.~\ref{fig:rhoH_multiT}(c,d)]. 
Though we can fit $\rho_\mathrm{H}(H)$ below 50~K to a two-band relationship with only two free parameters (see model details in Ref.~\cite{eguchi2019robust}), the best fits to $\rho_\mathrm{H}$ do not account for the field dependence of the longitudinal resistivity. Likewise, a multiband fit to the curved AFM-phase $\rho_\mathrm{H}$ cannot fit the Hall and longitudinal resistivity well simultaneously, so we assume that a Hall contribution emerges in the AFM phase that is not exclusively from the multiband physics observed between 20 and 50~K, leaving side-jump scattering and momentum-space Berry curvature as possible dominant contributions. Additional data features point to a significant intrinsic contribution associated with the AFM phase. First, the temperature dependence of $\rho_{zx}$ shows a sharp transition at $T_\mathrm{N}$ (Fig.~\ref{fig:rho_ca_cc_aa_T}). Second the high-field spin-reorientation phases below 20~K have a linear Hall resistivity with increasing field, following the linear-in-field magnetization and longitudinal resistivity, while the AFM phase shows a more complex field-dependent response. Third, an orientation-dependent zero-field Hall resistivity and remnant magnetization only present below $T_\mathrm{N}$ suggest the emergence of a contribution directly related to the symmetry of the AFM phase.

\subsection{Zero-field Hall resistivity and magnetic symmetry} \label{sec:zero_field}
\begin{figure*}
    \centering
    \subfloat{\includegraphics[width=0.815\columnwidth]{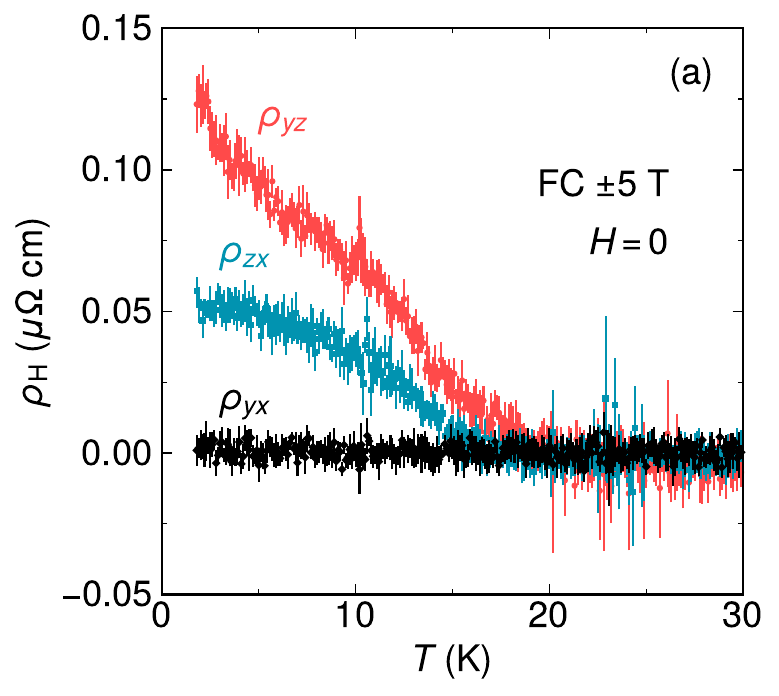}}
    \hspace{1.5em}
    \subfloat{\includegraphics[width=0.8\columnwidth]{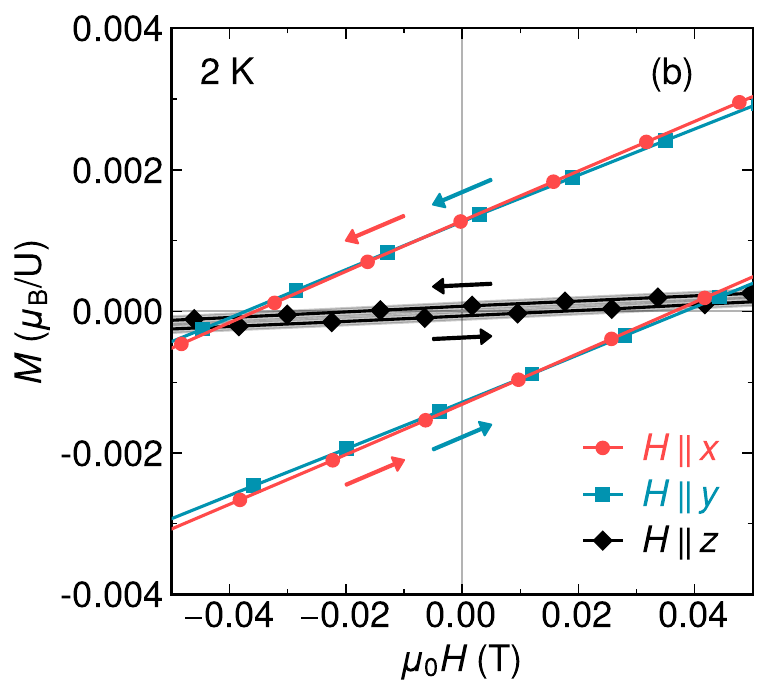}}
    \caption{(a) The zero-field Hall $\rho_\mathrm{H}^0$ of \mat\ as a function of temperature is shown for three different orientations. (b) The remnant magnetization of \mat\ at 2~K is shown with gray shading representing the instrument error combined with a potential systematic error of $\pm$2.5$^\circ$ sample misalignment relative to a magnetic field applied along $z$.}
    \label{fig:zerofield}
\end{figure*}

A hallmark of a momentum-space Berry curvature contribution to the Hall effect is an observed transverse resistivity in the magnetically ordered phase at zero applied magnetic field, where the ordinary contribution is zero. In \mat, after field cooling, a finite, temperature-dependent Hall signal at zero applied magnetic field is observed below $T_\mathrm{N}$ for $\rho_{yz}$ and $\rho_{zx}$, where it was concluded that the  Hall resistivity in the AFM phase stems from an intrinsic contribution, but not for $\rho_{yx}$, as displayed in  Fig.~\ref{fig:zerofield}(a). In contrast, no finite value is found above $T_\mathrm{N}$. The zero-field resistivity values are converted to conductivity at 2~K using Eq.~(\ref{eq:sig-H}), yielding $\sigma_{zy}=5.6(4)$~$\Omega^{-1}$~cm$^{-1}$ and $\sigma_{xz}=3.7(1)$~$\Omega^{-1}$~cm$^{-1}$. 
The zero-field Hall signal for two orientations cannot be attributed to a current-induced magnetization. For both $H{\parallel}x$ and $H{\parallel}y$, a remnant magnetization of $\sim$0.001~$\mu_\mathrm{B}$/U ($\mu_\mathrm{0}M{\approx}0.2$~mT) is observed upon reducing the field from 16~T [Fig.~\ref{fig:zerofield}(b)]. 
Any current-induced magnetization will be significantly smaller than the remnant magnetization since our current density (7.5~kA/m$^2$ for $\rho_{yz}$ and 6.0~kA/m$^2$ for $\rho_{zx}$) is smaller than that of Ref.~\cite{saito2018evidence} (40--60~kA/m$^2$), where the induced magnetization is only of order 10$^{-6}$~$\mu_\mathrm{B}$/U. 
In contrast to an in-$xy$-plane applied magnetic field, for $H{\parallel}z$, the remnant magnetization is small and is zero within the error for a sample misalignment of 3$^\circ$ [Fig.~\ref{fig:zerofield}(b)].
The zero-field Hall resistivity does not scale linearly with the remnant magnetization, as shown in Fig.~S3 \cite{supplemental}. 

\begin{figure}
    \centering
    \includegraphics[width=\columnwidth]{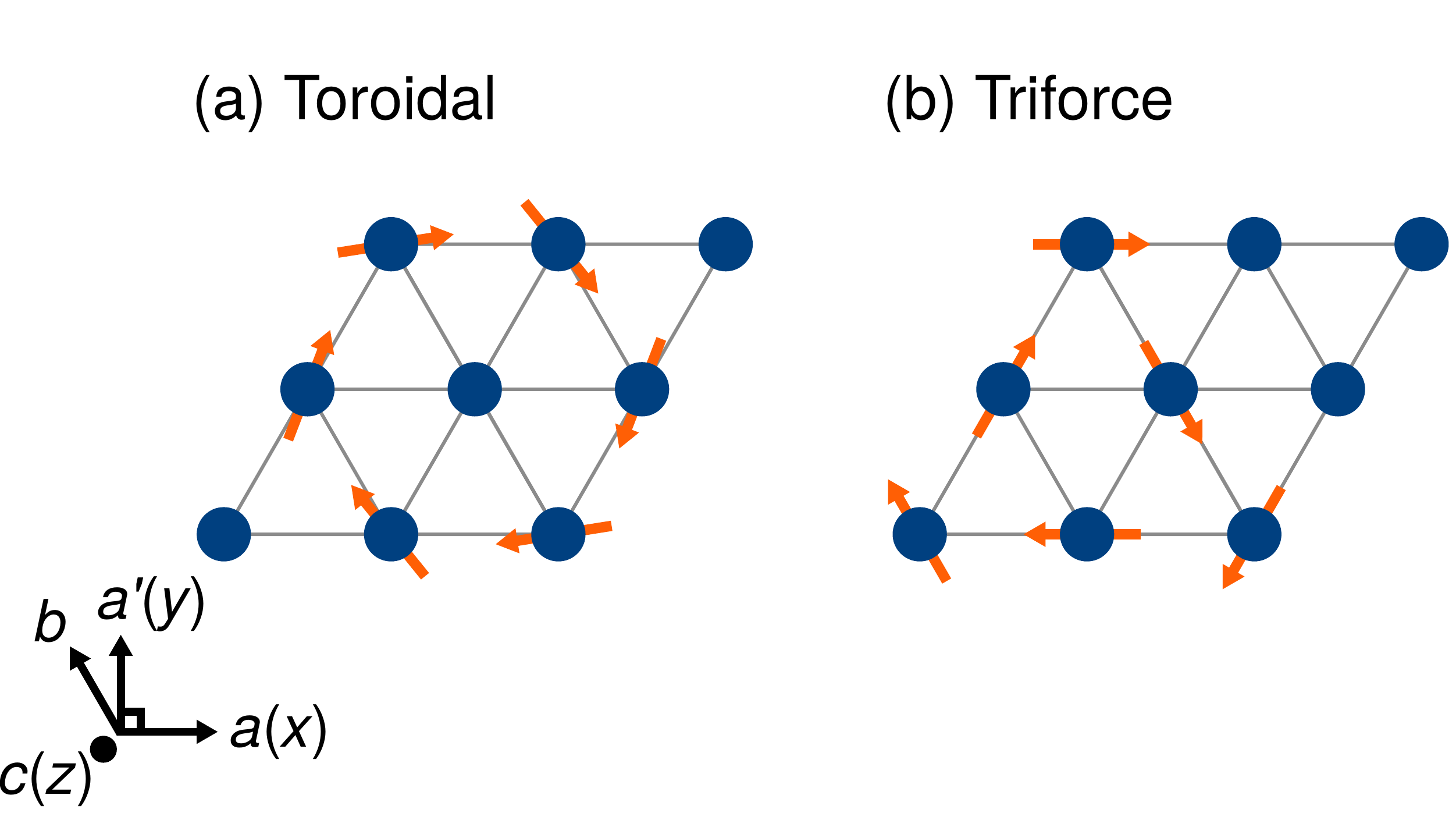}
    \caption{The basic units of the two magnetic ordering types proposed for \mat\ are shown with ordered uranium dipole moment directions based on Ref.~\cite{willwater2021crystallographic} for the (a) toroidal arrangement and on Ref.~\cite{ishitobi2023triple} for the (b) triforce one. Note that the two sets of uranium atoms in (a) and (b) are different sites in the full magnetic cells \cite{ishitobi2023triple}.}
    \label{fig:magorder}
\end{figure}

The observation of a zero-field Hall resistivity/conductivity is tied to the zero-field magnetic order of \mat. We assume that multipolar order, which has been proposed as the origin of nontrivial field-dependent Hall effects in, e.g., URhSn \cite{shimizu2026hall}, does not generate restrictions on the observed zero-field Hall effect and only consider the ordering of dipole magnetic moments in our symmetry analysis. The two primary proposed magnetic orders, toroidal and triforce, are shown in Fig.~\ref{fig:magorder}. For the toroidal magnetic order [Fig.~\ref{fig:magorder}(a)], magnetic moments are ordered on the $Cmcm$ $8f$ Wyckoff position uranium sites (``U2" and ``U3" in Ref.~\cite{willwater2021crystallographic}), which split into sites with $2g$ and $2h$ symmetry in the $Pmm2$ cell, but not the $Cmcm$ $4c$ sites (``U1" and ``U4" in Ref.~\cite{willwater2021crystallographic}). For the triforce magnetic order, some of the $Cmcm$ $4c$ sites are ordered, including sites in the $Pmm2$ cell at $1d$, $1b$, $2g$, and $2h$ Wyckoff positions, and some $Cmcm$ $8f$ sites are not ordered. Versions of both moment arrangements with modulated moment sizes have also been suggested in Ref.~\cite{ishitobi2023triple}, where the proposed magnetic space groups (MSGs) are $Pm'm2'$ for (un)modulated triforce order, $Pm'c2_1'$ for modulated toroidal order, and $Pmm'a$ for the (theoretical) single-$q$ order.
Notably, the $Pm'c2_1'$ and $Pmm'a$ MSGs are not magnetic subgroups consistent with the single propagation vector ($\textbf{k}$) magnetic structure of Ref.~\cite{willwater2021crystallographic} that has $Cmcm$ parent crystallographic symmetry and $\textbf{k}=(0,2/3,0)$ or, by extension, $Pmm2$ symmetry and $\textbf{k}=(0,0,2/3)$ [$b$ and $c$ axes switched compared with $Cmcm$]. In contrast, $Pm'm2'$ is a maximal MSG of $Pmm2$ in the case of $\textbf{k}=(0,0,2/3)$ \cite{perez2015symmetry}.

To evaluate the possible magnetic order associated with the observed zero-field Hall effect, we assume a commensurate magnetic structure with $\textbf{k}=(0,2/3,0)$ and $Cmcm$ nonmagnetic parent space group symmetry or $\textbf{k}=(0,0,2/3)$ and $Pmm2$ symmetry. Enumerating the magnetic subgroups with \textsc{k\_subgroupsmag} \cite{perez2015symmetry}, one finds 51 options for $Cmcm$ and 11 for $Pmm2$, with 8 and 4 maximal MSG options, respectively. Next, MSGs are filtered out if all zero-field Hall tensor components are zero due to symmetry restrictions; if they cannot host a magnetoelectric response \cite{hayami2018classification}, which was previously observed for \mat\ \cite{saito2018evidence}; or if they do not allow for ordered magnetic dipole moments within the $xy$ plane (checked with \textsc{magmodelize} \cite{perez2015symmetry}). The momentum-space Berry curvature is unaffected by translation symmetry and can be described using the magnetic point group (MPG) rather than the full magnetic space group \cite{suzuki2017cluster}. 
Three orthorhombic MPGs allow for zero-field Hall and magnetoelectric effects: $2'2'2$, $m'm2'$, and $m'm'2$. Note that all of the maximal MSGs of $Cmcm$ have incompatible MPG symmetry ($m'm'm'$, $m'm'm$, $m'mm$, or $mmm.1$). Of the orthorhombic subgroups, 10 for $Cmcm$ and 2 for $Pmm2$ have compatible MPG symmetry and allow in-$xy$-plane magnetic moment components. Each orthorhombic MPG, though, contains at least a two-fold rotation or mirror plane, both of which restrict two zero-field Hall components to zero, leaving only one component finite. The additional symmetry operations produce redundant restrictions on the Hall components. This raises the possibility that the observation of two finite zero-field Hall components stems from lower, monoclinic magnetic symmetry. Before lowering the symmetry to monoclinic, one must consider whether magnetic domains can produce the two observed finite zero-field Hall signals for $Cmcm$ or $Pmm2$ parent symmetry.

Even though a magnetic point group symmetry may only allow a single finite zero-field Hall tensor component, coexisting magnetic domains can lead to the observation of a finite zero-field Hall signal for multiple crystallographic orientations. For example, in Mn$_3$Sn, the room-temperature magnetic space group symmetry is $Cmc'm'$ (BNS \#63.463, n.b. BNS numbering is used throughout this work) \cite{cederholm2026ground}. Losing the six-fold symmetry of the parent space group $P6_3/mmc$ allows for magnetic domains to form where the mirror plane perpendicular to the $a$ axis of the $Cmc'm'$ moment arrangement can be rotated 120$^{\circ}$ with respect to the crystallographic parent $c$ axis. Multiple domains can, therefore, be populated after field cooling \cite{cederholm2026ground}. Consequently, a finite zero-field Hall effect is observed for two Hall orientations \cite{nakatsuji2015large} despite the $Cmc'm'$ symmetry only allowing for one finite component ($\rho_{bc}^0$ with respect to the $Cmc'm'$ axes) in a single magnetic domain \cite{suzuki2017cluster,gallego2019automatic}. For \mat, the domains and their rotational symmetry must be considered before evaluating the possible magnetic symmetries dictating the zero-field Hall effect.

We assume that the \mat single crystal sample is a single crystallographic domain based upon the Laue diffraction results and that our applied magnetic field history (maximum field only $\pm$5~T) for each Hall bar results in the same zero-field antiferromagnetic phase. The increase in the unit cell size implied by the experimental magnetic propagation vector and the loss of symmetry operations (including the invariance under time reversal of the nonmagnetic phase's ``gray" space group) caused by the transition to the ordered antiferromagnetic phase can lead to magnetic domain formation \cite{perez2015symmetry}. Field cooling under opposite magnetic fields is expected to select opposite time-reversal-related domain sets, thus preventing the cancellation of the momentum-space Berry curvature and, by extension, the zero-field Hall effect. This cancellation is why the zero-field cooled Hall data in Fig.~\ref{fig:panel} is zero at zero applied field. The magnetic domain possibilities not simply related by time-reversal are, therefore, relevant.

The principal axes of each orthorhombic magnetic subgroup are parallel to a principal ($x-y-z$) axis of the parent crystallographic space group, though they may change definition based on the unit-cell setting. For example, the $Am'a'2$ $a$ axis corresponds to the $Cmcm$ $c$ axis. 
Depending on the magnetic subgroup, magnetic domains for $Cmcm$ or $Pmm2$ parent symmetry can be related by two-fold rotations around a principal parent/crystallographic axis, mirror planes perpendicular to a parent/crystallographic axis, or an inversion center (only $Cmcm$) with respect to the parent/crystallographic axes. In some cases, these operations, or the identity operation, can be paired with translations related to a loss of $C$ centering for $Cmcm$ or to the expansion of the unit cell, which generates multiple origin options with respect to the parent axes. Unlike the three-fold rotation of magnetic domains in Mn$_3$Sn, none of these operations changes the orientation of symmetry-restricting elements in the magnetic subgroups (mirrors and rotations) with respect to the parent axes in \mat. To confirm that each magnetic domain produces the same symmetry restriction(s) with respect to the parent axes, we derived the magnetic domain possibilities using \textsc{mvisualize} \cite{perez2015symmetry} and submitted the resulting domain options to \textsc{mtensor} \cite{gallego2019automatic} for every remaining magnetic subgroup. Representative examples are shown in Appendix~\ref{ap:domains} for the $Am'a'2$ (\#40.207) subgroup of $Cmcm$ and the $Pm'm2'$ (\#25.60) subgroup of $Pmm2$. Populating multiple orthorhombic magnetic domains, therefore, will not lead to two finite zero-field Hall components, and monoclinic MSGs must be considered. 

Following the same procedure for the orthorhombic MSGs, we evaluate the monoclinic subgroup options. Of the monoclinic MPGs without an inversion operation, two allow for a single zero-field Hall component ($2.1$ and $m.1$) and require domain effects to explain the observed two components, and two allow for two zero-field Hall components ($2'$ and $m'$) \cite{gallego2019automatic}. The symmetry element associated with the monoclinic $b$ axis restricts the zero-field Hall tensor components. Similar to the orthorhombic cases, the monoclinic $b$ axis for every possible magnetic subgroup is parallel to a principal axis of the parent space group, and the same types of symmetry operation combinations relating magnetic domains exist. As before, we confirmed for each subgroup option that the monoclinic MPGs $2.1$ and $m.1$ paired with domain effects cannot explain the observed zero-field Hall effect. We are left with 6 monoclinic magnetic subgroups for $Cmcm$ parent symmetry and 3 for $Pmm2$ parent symmetry. 
For both parent space group types, the $2'$ rotation axis must be parallel to the parent $z$ axis, or the $m'$ plane must be perpendicular to the parent $z$ axis to allow the observed zero-field Hall resistivity components ($\rho_{yz}^0$ and $\rho_{zx}^0$),
leaving options with $Cm'$, $C2'$, and $Pm'$ symmetry, which are cataloged in Table~\ref{tab:msg_options}. 

\begin{table*}
    \centering
    \setlength{\tabcolsep}{7pt}
    \caption{Magnetic space groups are presented that are compatible with neutron diffraction experiments \cite{mentink1994magnetic,willwater2021crystallographic} and have symmetry allowing the observed finite zero-field Hall components ($\rho_{yz}$, $\rho_{zx}$) and magnetoelectric effects. The transformation matrices relate the standard-setting unit-cell basis $\textbf{A}''=(\textbf{a}'',\textbf{b}'',\textbf{c}'')$ of the magnetic space group to the basis $\textbf{A}=(\textbf{a},\textbf{b},\textbf{c})$ of the nonmagnetic parent group by $\textbf{A}''=\textbf{A}\textbf{P}$ paired with a unit cell origin shift $\textbf{p}=(p_1\textbf{a},p_2\textbf{b},p_3\textbf{c}$).}
    \begin{tabular}{ccccc}
    \midrule
    Parent & MSG & Transformation & Allowed Moment & Allowed $\alpha^\mathrm{E/J}$ \\
    & & \textbf{P}, \textbf{p} & Components & \\
    \midrule
    $Cmcm$ & $Cm'$ (\#8.34)& 
    $\begin{pmatrix}
        0 & -1 & 0 \\
        3 & 0 & 0 \\
        0 & 0 & 1 \\
    \end{pmatrix}$,
    $\begin{pmatrix}
        0 & 0 & 0
    \end{pmatrix}$ 
    & ($M_x$,$M_y$,0) & \parbox{4cm}{\linespread{1.3}\selectfont$\alpha^\mathrm{E}_{yx}$, $\alpha^\mathrm{E}_{xy}$, $\alpha^\mathrm{E}_{xx}$, $\alpha^\mathrm{E}_{yy}$, $\alpha^\mathrm{E}_{zz}$,  \\ $\alpha^\mathrm{J}_{yz}$, $\alpha^\mathrm{J}_{zy}$, $\alpha^\mathrm{J}_{xz}$, $\alpha^\mathrm{J}_{zx}$} \\ \\
    $Cmcm$ & $C2'$ (\#5.15) &
        $\begin{pmatrix}
        0 & 1 & 0 \\
        -3 & 0 & 0 \\
        0 & 0 & 1 \\
    \end{pmatrix}$,
    $\begin{pmatrix}
        0 & \frac{1}{2} & 0
    \end{pmatrix}$
    & ($M_x$,$M_y$,$M_z$) & \parbox{4cm}{\linespread{1.3}\selectfont$\alpha^\mathrm{E}_{yz}$, $\alpha^\mathrm{E}_{zy}$, $\alpha^\mathrm{E}_{xz}$, $\alpha^\mathrm{E}_{zx}$, \\
    $\alpha^\mathrm{J}_{yx}$, $\alpha^\mathrm{J}_{xy}$, $\alpha^\mathrm{J}_{xx}$, $\alpha^\mathrm{J}_{yy}$, $\alpha^\mathrm{J}_{zz}$}\\ \\
    $Pmm2$ & $Pm'$ (\#6.20) &
    $\begin{pmatrix}
        0 & 1 & 0 \\
        -1 & 0 & 0 \\
        0 & 0 & 3 \\
    \end{pmatrix}$,
    $\begin{pmatrix}
        0 & 0 & 0
    \end{pmatrix}$
    & ($M_x$,$M_y$,0) & \parbox{4cm}{\linespread{1.3}\selectfont$\alpha^\mathrm{E}_{yx}$, $\alpha^\mathrm{E}_{xy}$, $\alpha^\mathrm{E}_{xx}$, $\alpha^\mathrm{E}_{yy}$, $\alpha^\mathrm{E}_{zz}$,  \\ $\alpha^\mathrm{J}_{yz}$, $\alpha^\mathrm{J}_{zy}$, $\alpha^\mathrm{J}_{xz}$, $\alpha^\mathrm{J}_{zx}$}\\
    \midrule
    \end{tabular}
    \label{tab:msg_options}
\end{table*}

For each option, the symmetry is further analyzed on the basis of the previously reported magnetoelectric response of \mat\ \cite{saito2018evidence}. The authors of Ref.~\cite{ishitobi2023triple} show that the triforce order with $Pm'm2'$ symmetry accounts for the two reported magnetoelectric coefficients, $\alpha_{yx}$ and $\alpha_{yz}$ ($M_i=\alpha_{ij}E_j$), through an electric-field-driven contribution ($\alpha^\mathrm{E}$) and through a current-driven contribution ($\alpha^\mathrm{J}$), respectively. The electric-field-driven effect is an equilibrium effect associated with magnetic(-toroidal) multipoles and interband contributions, while the current-driven one is a transport effect associated with electric(-toroidal) multipoles and intraband contributions (of similar origin to the spin-Edelstein effect \cite{edelstein1990spin}) \cite{watanabe2018group,hayami2018classification}. The symmetry of the electric-field-driven response tensor in Jahn symmetry notation is aeV$^2$ (i.e., a time-reversal-odd, second-rank axial tensor with no (anti)symmetric constraints on $\alpha_{ij}-\alpha_{ji}$ pairs \cite{jahn1949note,gallego2019automatic}). Using $\textsc{mtensor}$, one can determine that the electric-field-driven effect may produce a finite $\alpha_{yx}^\mathrm{E}$ for $Cm'$, $\alpha_{yz}^\mathrm{E}$ for $C2'$, and $\alpha_{yx}^\mathrm{E}$ for $Pm'$. The current-driven response tensor follows the Jahn symmetry symbol for the Edelstein response tensor (eV$^2$, a time-reversal-even, second-rank rank axial tensor \cite{tenzin2023analogs}). We find that the electric-current-driven effect may produce a finite $\alpha_{yz}^\mathrm{J}$ for $Cm'$, $\alpha_{yx}^\mathrm{J}$ for $C2'$, and $\alpha_{yz}^\mathrm{J}$ for $Pm'$. Combining the results for the two origins of a magnetoelectric response, the $Cm'$, $C2'$, and $Pm'$ MSGs may account for the observed magnetoelectric effects. Additional allowed magnetoelectric effect components are listed in Table~\ref{tab:msg_options}.

Finally, for all three options, the allowed magnetic moment directions apply to any uranium site, and, thus, all three MSGs can host the proposed toroidal or triforce order from Refs.~\cite{mentink1994magnetic,willwater2021crystallographic,ishitobi2023triple}. For toroidal order, we note that $Cm'$ and $Pm'$ allow for a magnetic toroidal moment (a polar, time-reversal-odd vector) along the $z$ axis because the $m'$ planes are perpendicular to the $z$ axis. In contrast, $C2'$ only allows for toroidal moment components within the $xy$ plane because the $2'$ rotation axis is the $z$ axis. The $C2'$ symmetry, therefore, conflicts with neutron diffraction experiments \cite{mentink1994magnetic,willwater2021crystallographic} that are consistent with an in-plane ($xy$) U moment arrangement, which would produce an out-of-plane ($z$) toroidal moment. $Cm'$ and $Pm'$, moreover, restrict the ordered magnetic dipole moments to the $xy$ plane while $C2'$ allows an unobserved out-of-plane ($z$) component.

Both the $Cm'$ and $Pm'$ symmetry magnetic unit cells derived for the $Cmcm$ and $Pmm2$ parent space groups, respectively, may account for the observation of ordered magnetic dipoles on specific uranium sites, an arrangement of magnetic moments restricted to the $xy$ plane producing an out-of-plane ($z$) toroidal moment, the propagation vector $\textbf{k}=(0,2/3,0)$ [$Cmcm$] or $\textbf{k}=(0,0,2/3)$ [$Pmm2$], two finite zero-field Hall effect tensor components, and the previously noted magnetoelectric effects for \mat. Therefore, we consider them strong possibilities for the zero-field magnetic ground state symmetry of \mat. We encourage further examination of the zero-field magnetic structure with highly sensitive probes of magnetic order, e.g., resonant inelastic X-ray scattering and nuclear magnetic resonance. A summary of the magnetic subgroup options and the primary reason for their incompatibility with experimental data are listed in the Supplemental Material \cite{supplemental}.

\subsection{Field-dependent anomalous Hall conductivity}
Having established with the finite zero-field Hall resistivity that a momentum-space Berry curvature contribution is present in the AFM phase, we evaluate the field dependence of this contribution by subtracting an ordinary Hall term using an estimated ordinary Hall coefficient ($R_\mathrm{0}$). The following estimates assume that there is negligible field dependence of $R_\mathrm{0}$ below $T_\mathrm{N}$.
Frequently, $R_\mathrm{0}$ is extracted from the field-polarized state where the ordinary Hall contribution presumably dominates, but the saturation field of \mat\ is not reached below 50~T \cite{mentink1995magnetization,mentink1996uni4b}. 
Therefore, we approximate $R_0$ using $\rho_{yx}(H{\parallel}z)$ below 7~T, where the zero-field Hall resistivity indicates no intrinsic/topological contribution. The resulting $R_\mathrm{0}$ is $0.23~\mu\Omega$~cm~T$^{-1}$ ($=0.0023$~cm$^3$/C), which is reasonable based on the paramagnetic region $R_\mathrm{0}$ values in the Supplemental Material \cite{supplemental}. 
After subtracting the ordinary Hall estimate, we convert the residual Hall resistivity ($\Delta\rho_\mathrm{H} = \rho_\mathrm{H}-R_\mathrm{0}\mu_\mathrm{0}H$) to conductivity ($\Delta\sigma_\mathrm{H}$) using Eq.~(\ref{eq:sig-H}), replacing $\rho_{ji}$ in the numerator with $\Delta\rho_\mathrm{H}$ (Fig.~\ref{fig:hall_sigma}). For reference, $\sigma_{zy}$ and $\sigma_{xz}$ are provided in the Supplemental Material without an ordinary Hall term subtracted \cite{supplemental}. The maxima of $\Delta\sigma_\mathrm{H}$ are 89~$\Omega^{-1}$~cm$^{-1}$ for $\sigma_{zy}$ and 142~$\Omega^{-1}$~cm$^{-1}$ for $\sigma_{xz}$. An alternative approach to estimating $R_\mathrm{0}$ fits $\rho_\mathrm{H}$ just above $T_\mathrm{N}$ at 25~K as $\mu_\mathrm{0}H\rightarrow0$, i.e. as $\mu_\mathrm{0}M\rightarrow0$, for data collected after zero-field cooling. In this case, $R_\mathrm{0}=0.351~\mu\Omega$~cm~T$^{-1}$ for $\rho_{yz}(H{\parallel}x)$, and $R_\mathrm{0}=0.406~\mu\Omega$~cm~T$^{-1}$ for $\rho_{zx}(H{\parallel}y)$. Naturally, the larger $R_\mathrm{0}$ values also lead to smaller $\Delta\sigma_\mathrm{H}$ maxima of 71~$\Omega^{-1}$~cm$^{-1}$ for $\sigma_{zy}$ and 100~$\Omega^{-1}$~cm$^{-1}$ for $\sigma_{xz}$.
While the zero-field Hall conductivity of \mat\ is relatively small, these field-dependent values are sizable and comparable to those observed in the $R$Mn$_6X_6$ family where the kagome lattice of Mn atoms generates magnetic frustration \cite{asaba2020anomalous,yin2020quantum,wang2021field}.
After passing through the first field-dependent magnetic transition, the sign of $\Delta\rho_\mathrm{H}$, and thus $\Delta\sigma_\mathrm{H}$, changes, suggesting the spin-reorientation transitions, which show a linear field dependence in the Hall response, alter the Fermi surface and thus the ordinary Hall contribution.

\begin{figure}
    \centering
    \includegraphics[width=0.85\columnwidth]{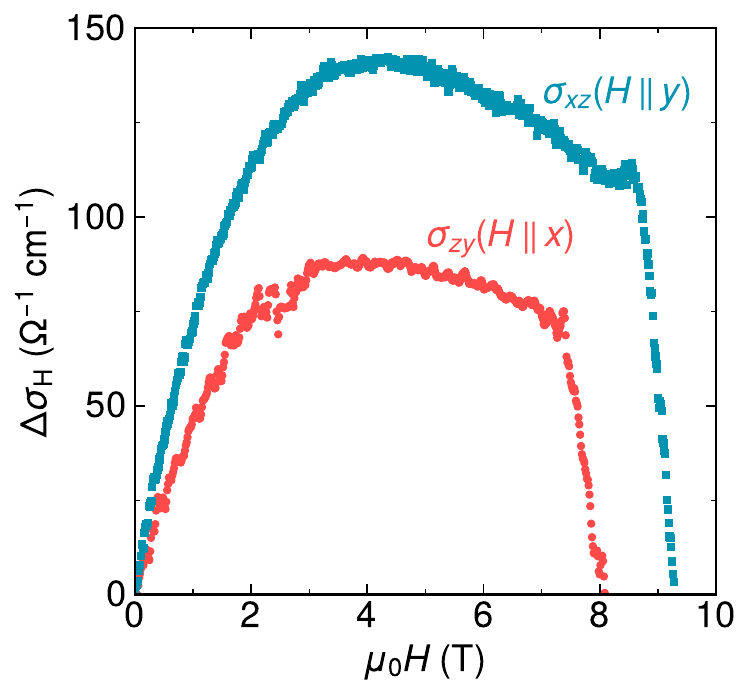}
    \caption{After subtracting an estimated ordinary Hall contribution, the residual conductivity is plotted. Since the data were collected after zero-field cooling, there is no zero-field signal due to randomized magnetic domains. The bump at 8.6~T for $\sigma_{xz}$ is due to a slight mismatch in the transition temperature for the longitudinal and Hall resistivity data used to convert to conductivity.}
    \label{fig:hall_sigma}
\end{figure}

\section{Conclusions}
In summary, the transverse and longitudinal resistivity, magnetic susceptibility, and magnetization are reported for \mat\ up to magnetic fields of 16 T. In the antiferromagnetic phase, the field-dependent Hall resistivity of \mat\ exhibits significant curvature vs. field when a magnetic field is applied within the $xy$ plane, but exhibits linear (trivial) behavior for a field applied along the $z$ axis. Analysis of the temperature and field dependence of the Hall effect shows that several contributions, including skew scattering, side-jump scattering, and the ordinary Hall effect, cannot fully explain this behavior. Consequently, on measuring the Hall effect at zero field as a function of temperature, a clear nonzero signal emerges for the orientations with a curved field dependence. Through a symmetry analysis, it is shown that a zero-field Hall signal for two unique orientations (finite $\rho_{yz}^0$ and $\rho_{zx}^0$) is incompatible with the previously proposed commensurate magnetic space groups. Considering the zero-field Hall effect, magnetic domain possibilities, and previous neutron diffraction and magnetoelectric effect experiments, we propose $Cm'$ or $Pm'$ as compatible magnetic subgroups for a structural $Cmcm$ or $Pmm2$ parent space group, respectively. The estimated maximum for the intrinsic anomalous Hall effect in the AFM phase is 89~$\Omega^{-1}$~cm$^{-1}$ for $\sigma_{zy}$ and 142~$\Omega^{-1}$~cm$^{-1}$ for $\sigma_{xz}$. This work adds to the growing literature indicating the magnetic ground state of \mat\ is nontrivial and deserves further experimental and theoretical exploration.

\section{Acknowledgments}
The authors are grateful for helpful conversations with S. Z. Lin and Y. Matsuda. Work at Los Alamos National Laboratory was performed under the auspices of the U.S. Department of Energy, Office of Basic Energy Sciences, Division of Materials Science and Engineering and was supported by the project ``Quantum Fluctuations in Narrow Band Systems."

\appendix
\counterwithin{table}{section}
\section{Magnetic domain examples}
\label{ap:domains}
The possible magnetic domains of the MSGs $Am'a'2$ ($Cmcm$ parent) and $Pm'm2'$ ($Pmm2$ parent) are enumerated in Tables~\ref{tab:domains1} and \ref{tab:domains2}, respectively. ``Operations" are the coset representatives relating domains to the first one listed. Operations are expressed in Seitz notation \cite{glazer2014seitz,perez2015symmetry} and in the parent cell setting. For example, $\{2_\mathrm{010}~\mid~0~0~1/2\}$ is a two-fold rotation around the parent $b$ axis paired with a translation of 1/2 along the parent $c$ axis. Operations are grouped if they generate the same transformation matrix, which follows the conventions in the main text. The allowed zero-field Hall component for each domain is listed in parent ($a$,$b$,$c$) and ``unified" ($x$,$y$,$z$) notation.

\begin{table}[ht]
    \centering
    \setlength{\tabcolsep}{5pt}
    \caption{$Am'a'2$ (BNS \#40.207)}
    \begin{tabular}{ccc}
    \midrule
    Operations & Transformation & Allowed \\
    & \textbf{P}, \textbf{p} & $\rho_\mathrm{H}^0$ \\
    \midrule 
    
    \parbox{2.5cm}{\linespread{1.3}\selectfont $\{1~\mid~0\}$ \\ $\{2_\mathrm{010}~\mid~0~0~\frac{1}{2}\}$} & 
    $\begin{pmatrix}
        0 & 0 & -1 \\
        0 & 3 & 0 \\
        1 & 0 & 0 \\
    \end{pmatrix}$,
    $\begin{pmatrix}
        0 & \frac{1}{2} & 0
    \end{pmatrix}$ & \parbox{1.5cm}{\linespread{1.3}\selectfont$\rho_{bc}$ \\ $[\rho_{yx}]$} \\ \\
    
    \parbox{2.5cm}{\linespread{1.3}\selectfont$\{2_\mathrm{001}~\mid~0~0~\frac{1}{2}\}$ \strut \\ $\{1~\mid~0~1~0\}$} &
        $\begin{pmatrix}
        0 & 0 & 1 \\
        0 & -3 & 0 \\
        1 & 0 & 0 \\
    \end{pmatrix}$,
    $\begin{pmatrix}
        0 & \frac{5}{2} & \frac{1}{2}
    \end{pmatrix}$ & \parbox{1.5cm}{\linespread{1.3}\selectfont $\rho_{bc}$ \\ $[\rho_{yx}]$} \\ \\
    
    \parbox{2.5cm}{\linespread{1.3}\selectfont$\{2_\mathrm{001}~\mid~\frac{1}{2}~\frac{1}{2}~\frac{1}{2}\}$ \\ $\{1~\mid~\frac{1}{2}~\frac{1}{2}~0\}$} &
        $\begin{pmatrix}
        0 & 0 & 1 \\
        0 & -3 & 0 \\
        1 & 0 & 0 \\
    \end{pmatrix}$,
    $\begin{pmatrix}
        \frac{1}{2} & 0 & \frac{1}{2}
    \end{pmatrix}$ & \parbox{1.5cm}{\linespread{1.3}\selectfont$\rho_{bc}$ \\ $[\rho_{yx}]$} \\ \\
        
    \parbox{2.5cm}{\linespread{1.3}\selectfont $\{1'~\mid~0\}$ \\ $\{2'_\mathrm{010}~\mid~0~0~\frac{1}{2}\}$} & 
    $\begin{pmatrix}
        0 & 0 & -1 \\
        0 & 3 & 0 \\
        1 & 0 & 0 \\
    \end{pmatrix}'$,
    $\begin{pmatrix}
        0 & \frac{1}{2} & 0
    \end{pmatrix}$ & \parbox{1.5cm}{\linespread{1.3}\selectfont$\rho_{bc}$ \\ $[\rho_{yx}]$} \\ \\
    
    \parbox{2.5cm}{\linespread{1.3}\selectfont$\{2'_\mathrm{001}~\mid~0~0~\frac{1}{2}\}$ \strut \\ $\{1'~\mid~0~1~0\}$} &
        $\begin{pmatrix}
        0 & 0 & 1 \\
        0 & -3 & 0 \\
        1 & 0 & 0 \\
    \end{pmatrix}'$,
    $\begin{pmatrix}
        0 & \frac{5}{2} & \frac{1}{2}
    \end{pmatrix}$ & \parbox{1.5cm}{\linespread{1.3}\selectfont $\rho_{bc}$ \\ $[\rho_{yx}]$} \\ \\
    
    \parbox{2.5cm}{\linespread{1.3}\selectfont$\{2'_\mathrm{001}~\mid~\frac{1}{2}~\frac{1}{2}~\frac{1}{2}\}$ \\ $\{1'~\mid~\frac{1}{2}~\frac{1}{2}~0\}$} &
        $\begin{pmatrix}
        0 & 0 & 1 \\
        0 & -3 & 0 \\
        1 & 0 & 0 \\
    \end{pmatrix}'$,
    $\begin{pmatrix}
        \frac{1}{2} & 0 & \frac{1}{2}
    \end{pmatrix}$ & \parbox{1.5cm}{\linespread{1.3}\selectfont$\rho_{bc}$ \\ $[\rho_{yx}]$} \\
    \midrule
    \end{tabular}
    \label{tab:domains1}
\end{table}

\begin{table}[ht]
    \centering
    \setlength{\tabcolsep}{5pt}
    \caption{$Pm'm2'$ (BNS \#25.59)}
    \begin{tabular}{ccc}
    \midrule
    Operations & Transformation & Allowed \\
    & \textbf{P},\textbf{p} & $\rho_\mathrm{H}^0$ \\
    \midrule 
    
    \parbox{2.5cm}{\linespread{1.3}\selectfont $\{1~\mid~0\}$ \\ $\{1~\mid~0~0~1\}$ \\ $\{1~\mid~0~0~2\}$} & 
    $\begin{pmatrix}
        1 & 0 & 0 \\
        0 & 1 & 0 \\
        0 & 0 & 3 \\
    \end{pmatrix}$,
    $\begin{pmatrix}
        0 & 0 & 0
    \end{pmatrix}$ & \parbox{1.5cm}{\linespread{1.3}\selectfont$\rho_{ac}$ \\ $[\rho_{zy}]$} \\ \\

        \parbox{2.5cm}{\linespread{1.3}\selectfont $\{1'~\mid~0\}$ \\ $\{1'~\mid~0~0~1\}$ \\ $\{1'~\mid~0~0~2\}$} & 
    $\begin{pmatrix}
        1 & 0 & 0 \\
        0 & 1 & 0 \\
        0 & 0 & 3 \\
    \end{pmatrix}'$,
    $\begin{pmatrix}
        0 & 0 & 0
    \end{pmatrix}$ & \parbox{1.5cm}{\linespread{1.3}\selectfont$\rho_{ac}$ \\ $[\rho_{zy}]$} \\
    \midrule
    \end{tabular}
    \label{tab:domains2}
\end{table}

\bibliography{UNi4B.bib}

\end{document}


\begin{center}
\Large 
\textbf{Magnetic symmetry implications of the zero- and applied-field Hall effect of \mat}\\
\vspace{1em}
Supplemental Material\\
\vspace{1em}
\normalsize
Z.~W.~Riedel, W.~S.~Simeth, S.~M.~Thomas, F.~Ronning, and E.~D.~Bauer 
\end{center}

\vspace{0.5em}

\section{Additional magnetization data}
\begin{figure}[h]
    \centering
    \includegraphics[width=0.65\columnwidth]{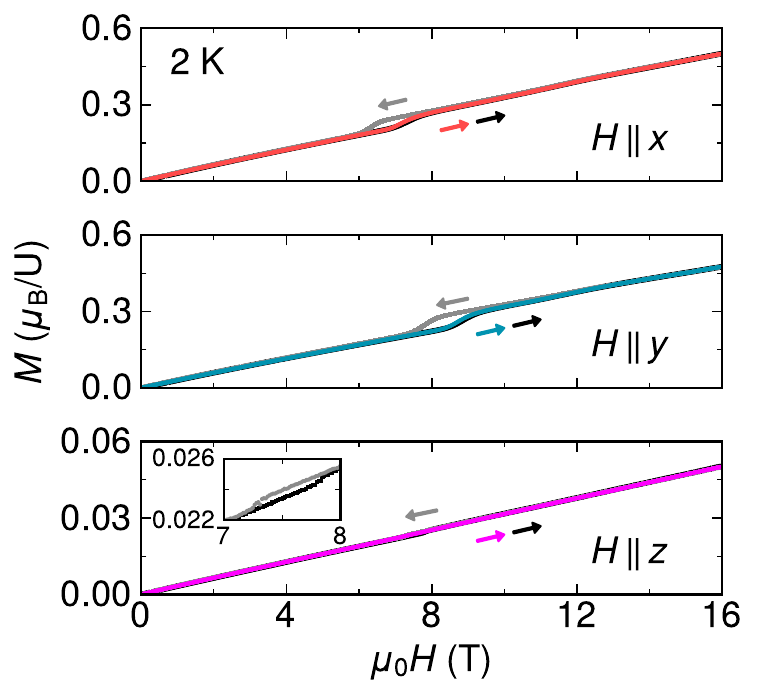}
    \caption{Magnetization as a function of applied field is plotted for the three principal axes of \mat\ at 2~K. Data are collected after zero-field cooling (colorful), decreasing field from 16~T (gray), and increasing field from -16~T (black). The zero-field cooling and increasing field sets are nearly indistinguishable for each orientation. (Inset) A previously unreported hysteretic transition appears around 7.5~T for $H{\parallel}z$.}
    \label{fig:M_xtraT}
\end{figure}

\begin{figure*}[h]
    \centering
    \subfloat{\includegraphics[width=0.45\columnwidth]{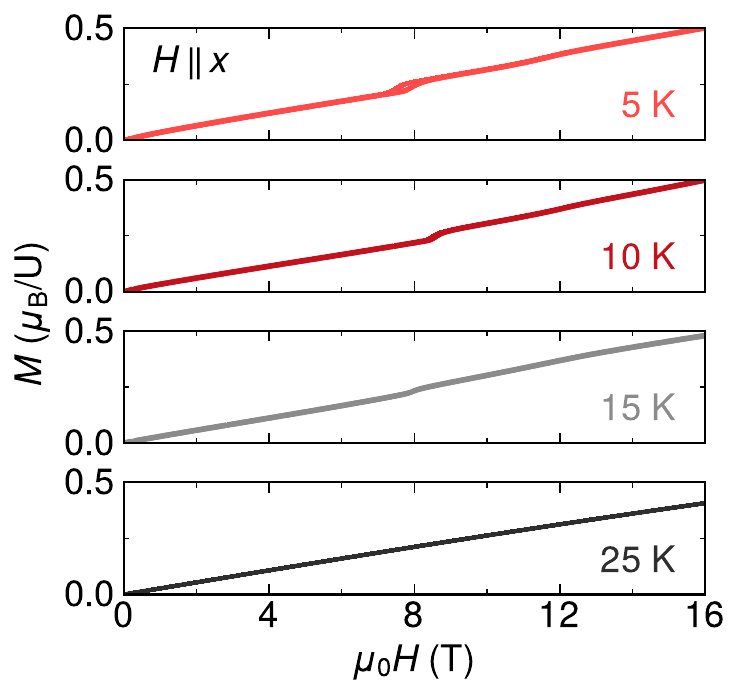}}
    \hspace{2em}
    \subfloat{\includegraphics[width=0.45\columnwidth]{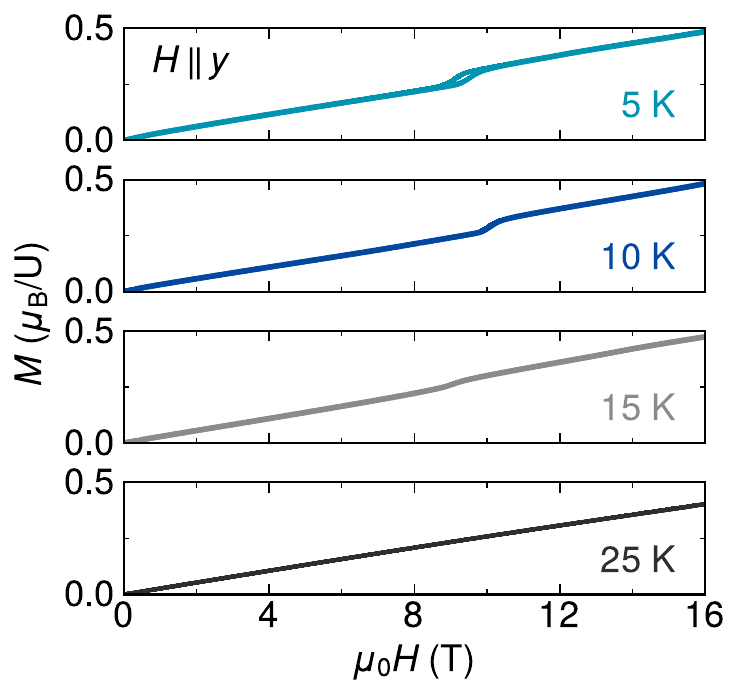}}
    \caption{Magnetization data for an in-$xy$-plane applied magnetic field shows little hysteresis above 2~K for either field orientation.}
    \label{fig:M_5to25K}
\end{figure*}

\begin{figure*}[h]
    \centering
    \subfloat{\includegraphics[width=0.45\columnwidth]{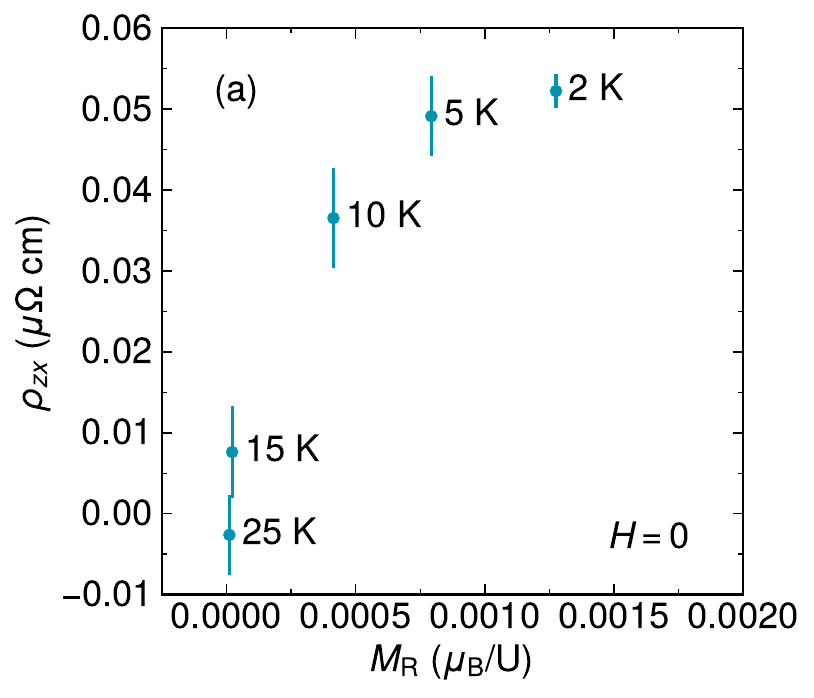}}
    \hspace{2em}
    \subfloat{\includegraphics[width=0.42\columnwidth]{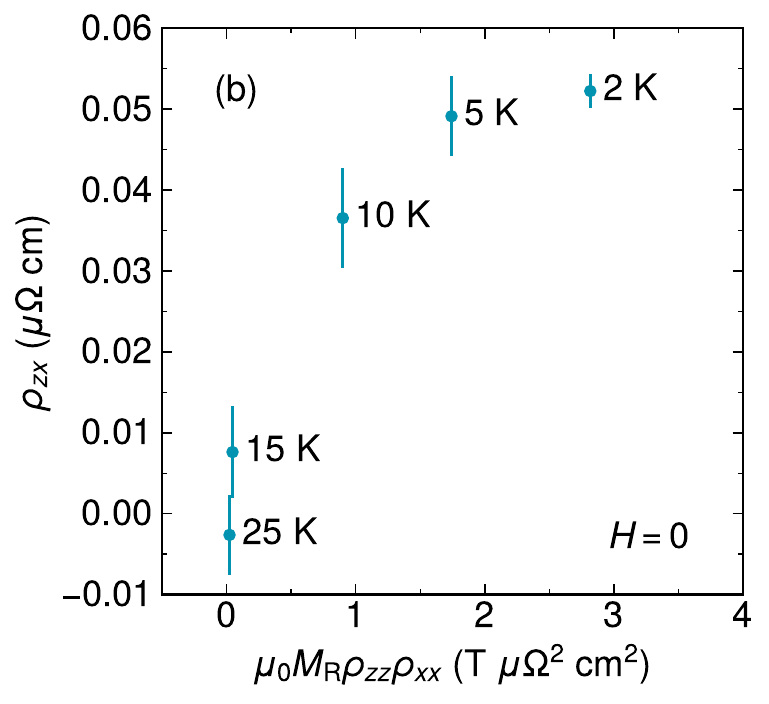}}
    \caption{The zero-field Hall resistivity does not scale with (a) the remnant magnetization [$M_\mathrm{R}=M(H=0)$ after approaching zero field from $\pm$16~T] or (b) $M_\mathrm{R}$ times the zero-field longitudinal resistivity.}
    \label{fig:remnantM}
\end{figure*}

Field-dependent magnetization data are presented for \mat\ at 2~K for each applied magnetic field direction in Fig.~\ref{fig:M_xtraT}. Data for 5--25~K and applied fields within the $xy$ plane are in Fig.~\ref{fig:M_5to25K}. Further, Figure~\ref{fig:remnantM} shows that the zero-field Hall resistivity does not scale with the remnant magnetization, suggesting an intrinsic Berry curvature origin.

\clearpage
\section{Additional longitudinal resistivity/conductivity data}
\begin{figure}[h]
    \centering
    \subfloat{\includegraphics[width=0.45\columnwidth]{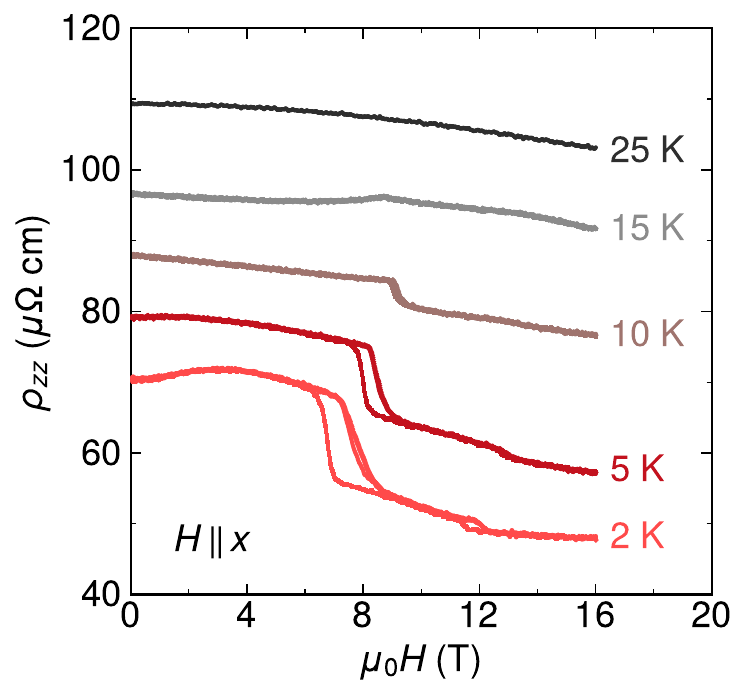}}
    \hspace{2em}
    \subfloat{\includegraphics[width=0.45\columnwidth]{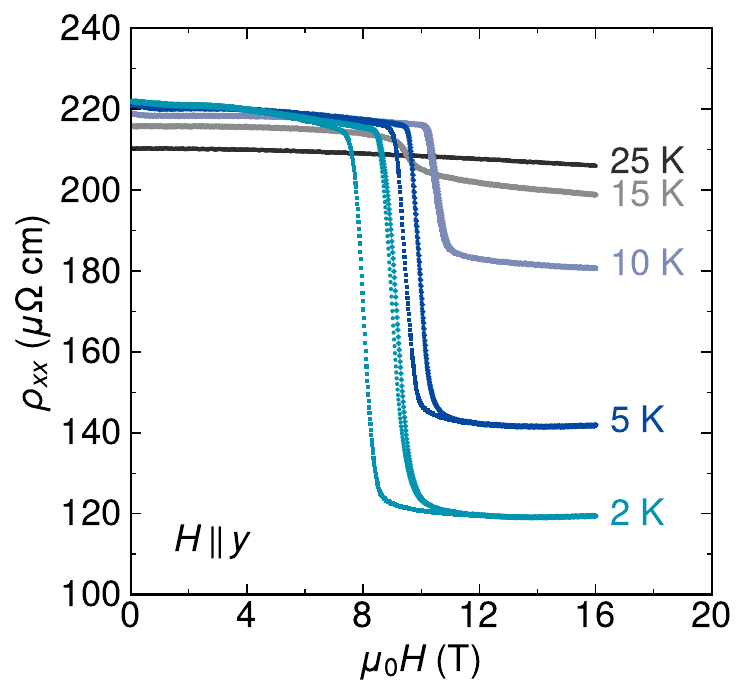}}
    \caption{The field-dependent longitudinal resistivity collected simultaneously with the Hall resistivity. Data were collected after zero-field cooling, on decreasing field from +16~T, and on increasing field from -16~T.}
    \label{fig:rhoxx_2to25}
\end{figure}

\begin{figure}[h]
    \centering
    \subfloat{\includegraphics[width=0.45\columnwidth]{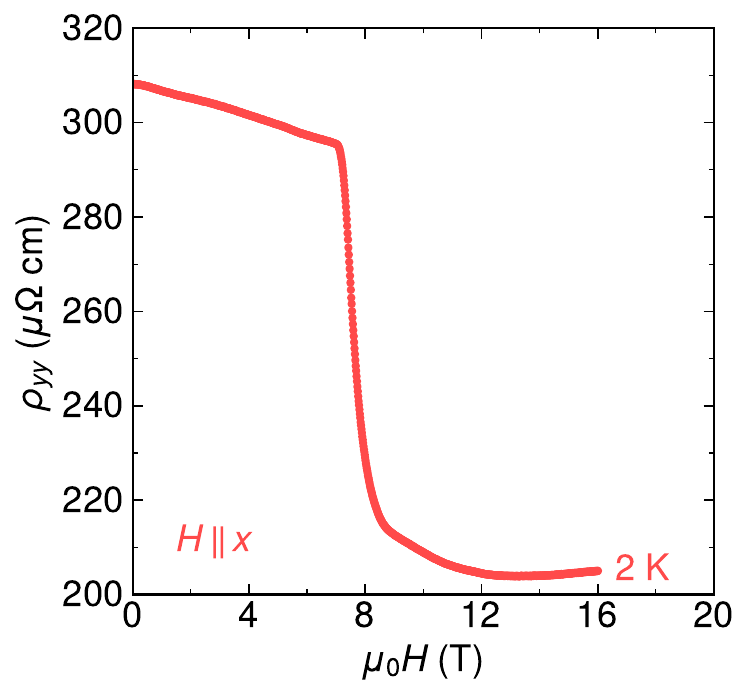}}
    \hspace{2em}
    \subfloat{\includegraphics[width=0.45\columnwidth]{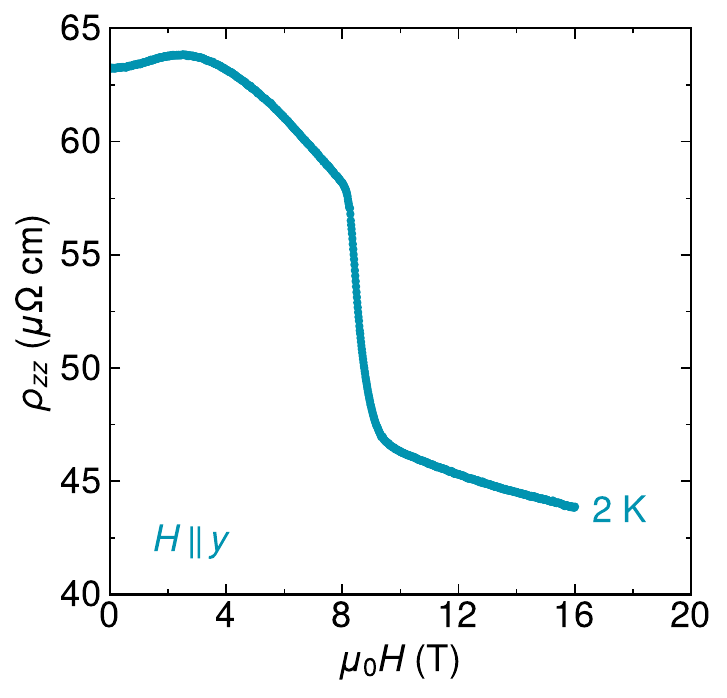}}
    \caption{Additional longitudinal resistivity datasets were collected for accurate conversion to conductivity at 2~K. Data were only collected after zero-field cooling.}
    \label{fig:rhoxx_for_sigma}
\end{figure}

\begin{figure}[h]
    \centering
    \subfloat{\includegraphics[width=0.45\columnwidth]{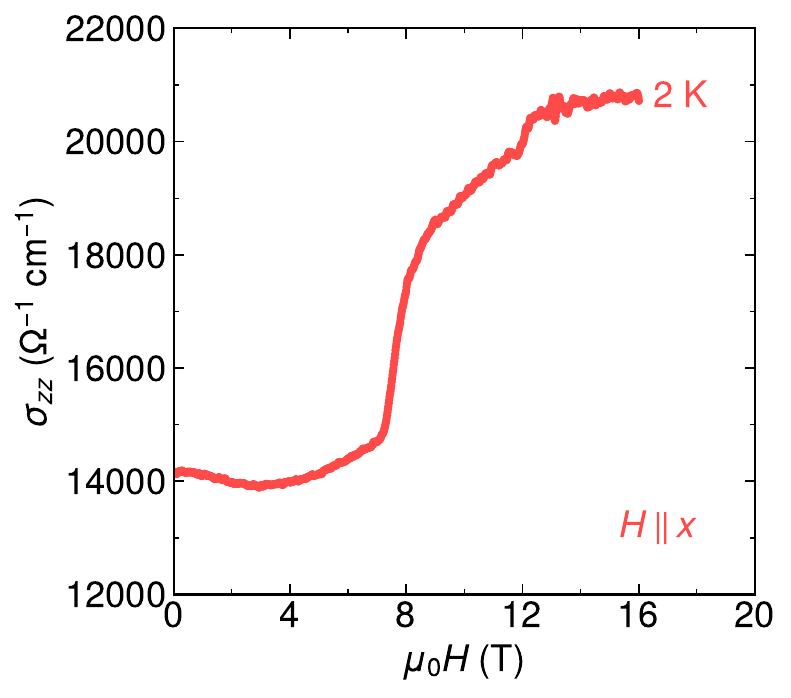}}
    \hspace{2em}
    \subfloat{\includegraphics[width=0.45\columnwidth]{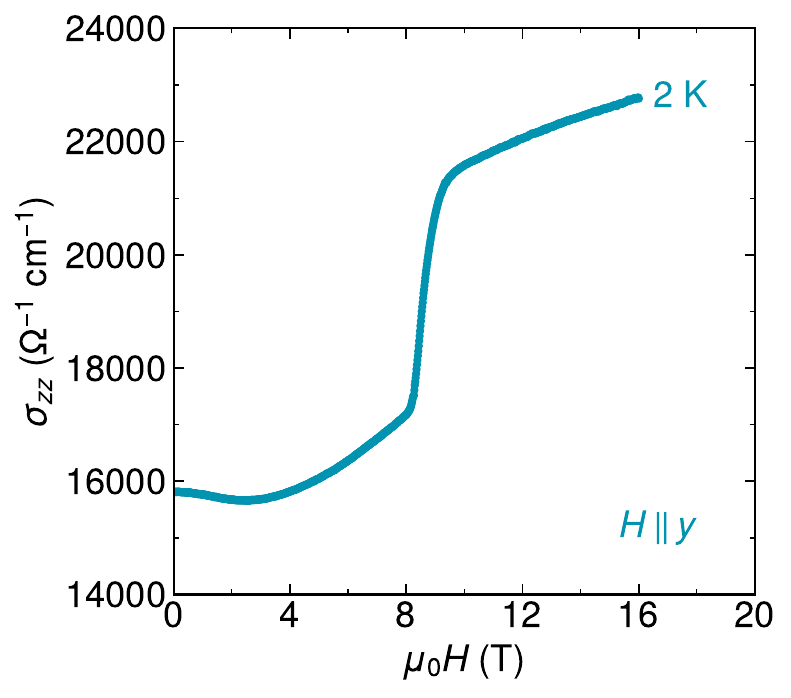}}
    \\   
    \subfloat{\includegraphics[width=0.45\columnwidth]{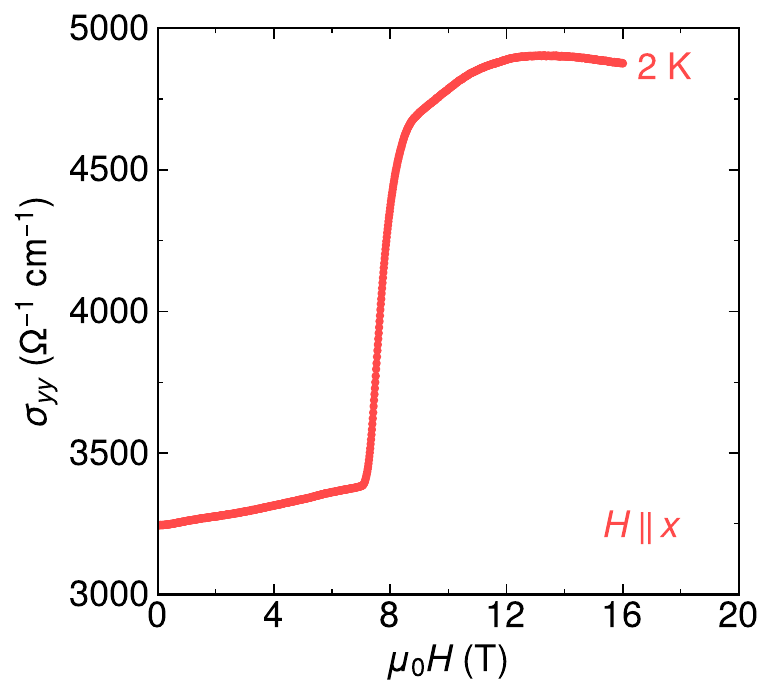}}  
    \hspace{2em}
    \subfloat{\includegraphics[width=0.45\columnwidth]{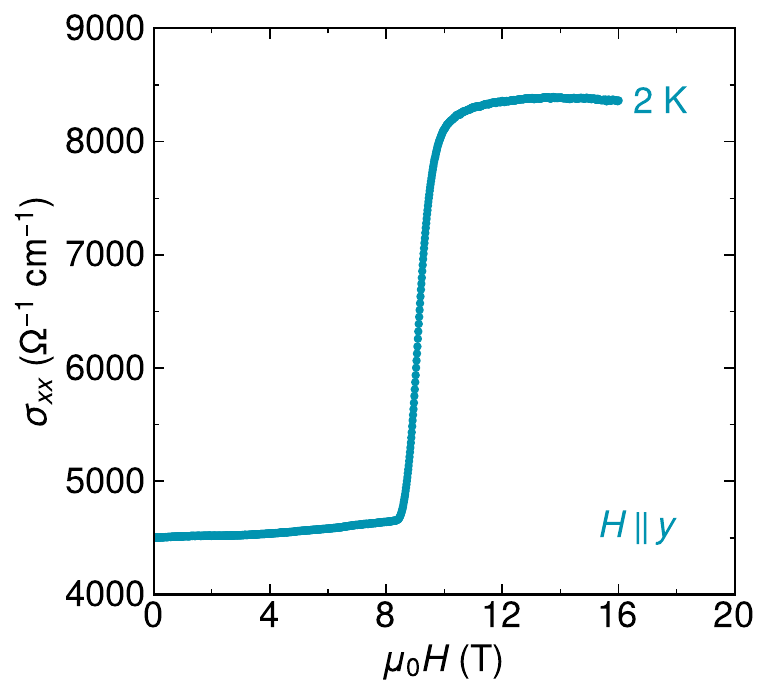}}
    \caption{The longitudinal conductivity is calculated for four orientations using the equation from the main text for $\sigma_{ii}$. Each orientation has a conductivity of order 10$^3$--10$^4$~$\Omega^{-1}$~cm$^{-1}$, placing \mat\ in the regime where intrinsic contributions to the Hall resistivity are expected to dominate.}
    \label{fig:longitudinal_conductivity}
\end{figure}

Longitudinal resistivity datasets for 2--25~K are presented in Figs.~\ref{fig:rhoxx_2to25} and \ref{fig:rhoxx_for_sigma} to show the evolution of the magnetic phase transitions with an applied field within the $xy$ plane and to convert to conductivity. Several 2~K data sets are converted to conductivity in Fig.~\ref{fig:longitudinal_conductivity} to show that the values are not in the region where skew scattering is expected to dominate ($\sigma_{ii}>10^6$~$\Omega^{-1}$~cm$^{-1}$).

\clearpage
\section{Additional Hall data and analysis}
\begin{figure}[h]
    \centering
    \includegraphics[width=0.55\columnwidth]{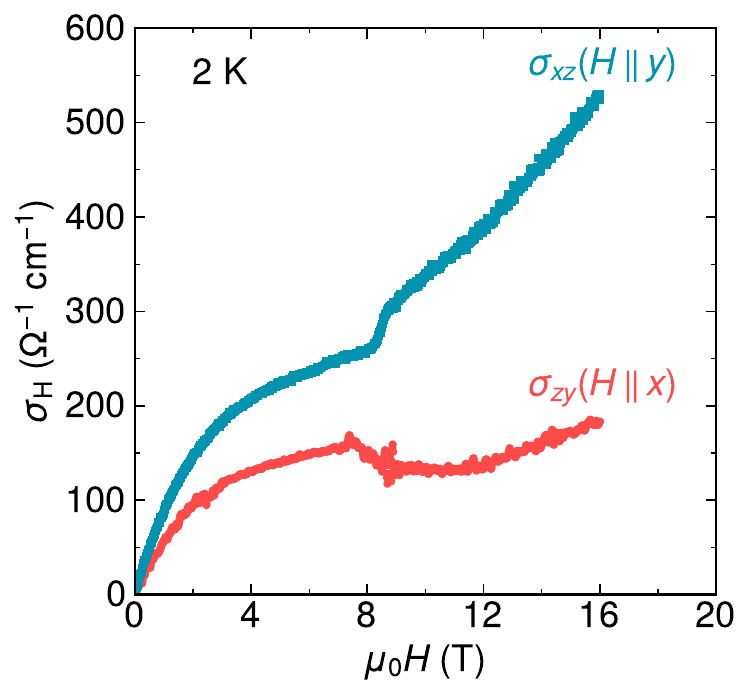}
    \caption{The field-dependent Hall conductivity is plotted for a magnetic field applied in the $xy$ plane.}
    \label{fig:sigH_2K}
\end{figure}

The 2~K Hall conductivity without any contributions subtracted is shown in Fig.~\ref{fig:sigH_2K}.

\clearpage
Temperature-implicit scaling of $\sigma_{xz}$ as a function of $\sigma_{xx}$ shows the inapplicability of a dominant skew scattering for the AFM phase. While $\sigma_{xz}$ is approximately linear as a function of $\sigma_{xx}$ above $T_\mathrm{N}$ (paramagnetic, PM), no linear relationship exists in the antiferromagnetic (AFM) phase.

\begin{figure}[h]
    \centering
    \includegraphics[width=0.55\columnwidth]{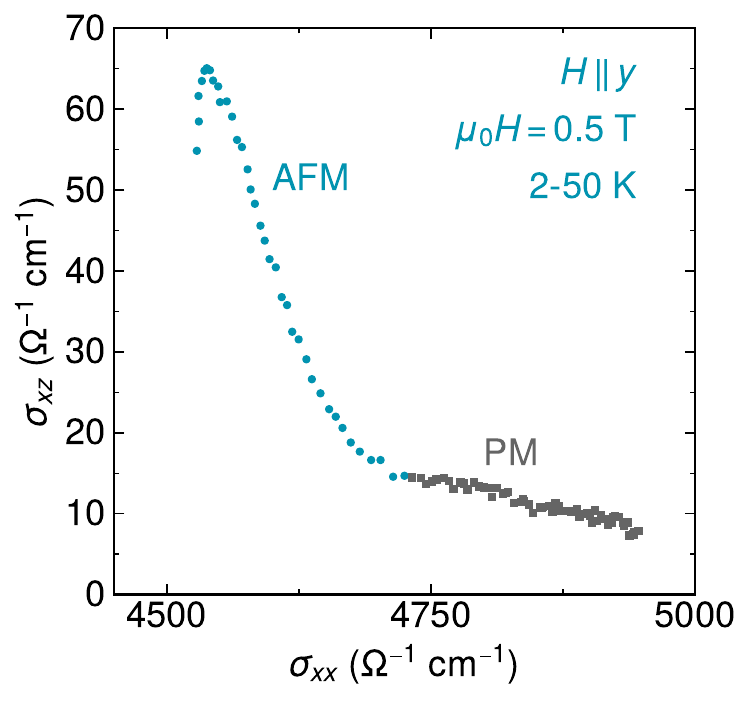}
    \caption{The Hall conductivity is scaled by the longitudinal conductivity using the empirical relationship for skew-scattering contributions. Each data point represents a unique temperature.}
    \label{fig:skew_scaling}
\end{figure}

\clearpage
The possible magnetic subgroups for commensurate, zero-field magnetic order of \mat\ are listed in Tables~\ref{tab:cmcm} and \ref{tab:pmm2} along with the primary reason for (in)compatibility with existing experimental data (more than one reason may exist). Transformation matrices relating the magnetic space group setting to the parent space group follow the same conventions as in the main text.

\renewcommand{\arraystretch}{0.78}
\begin{longtable}{llcc}
    \setlength{\tabcolsep}{3pt}\\

    \caption{Magnetic subgroups for $Cmcm$ with $\textbf{k}=(0,2/3,0)$.}
    \\ 
    \hline
    \\
    & MSG & Transformation & (In)compatibility \\
    & & \textbf{P}, \textbf{p} & \\ \\
    \hline
    \\
    1 & \parbox{1.8cm}{$Cm'c'm'$ \\ (\#63.465)} & 
    $\begin{pmatrix}
        1 & 0 & 0 \\
        0 & 3 & 0 \\
        0 & 0 & 1 \\
    \end{pmatrix}$,
    $\begin{pmatrix}
        0 & 1/2 & 0
    \end{pmatrix}$ & Incompatible MPG \\ \\
    2 & \parbox{1.8cm}{$Cm'cm'$ \\ (\#63.464)} & 
    $\begin{pmatrix}
        1 & 0 & 0 \\
        0 & 3 & 0 \\
        0 & 0 & 1 \\
    \end{pmatrix}$,
    $\begin{pmatrix}
        0 & 1/2 & 0
    \end{pmatrix}$ & Incompatible MPG \\ \\
    3 & \parbox{1.8cm}{$Cmc'm'$ \\ (\#63.463)} & 
    $\begin{pmatrix}
        1 & 0 & 0 \\
        0 & 3 & 0 \\
        0 & 0 & 1 \\
    \end{pmatrix}$,
    $\begin{pmatrix}
        0 & 1/2 & 0
    \end{pmatrix}$ & Incompatible MPG \\ \\
    4 & \parbox{1.8cm}{$Cm'c'm$ \\ (\#63.462)} & 
    $\begin{pmatrix}
        1 & 0 & 0 \\
        0 & 3 & 0 \\
        0 & 0 & 1 \\
    \end{pmatrix}$,
    $\begin{pmatrix}
        0 & 1/2 & 0
    \end{pmatrix}$ & Incompatible MPG \\ \\
    5 & \parbox{1.8cm}{$Cmcm'$ \\ (\#63.461)} & 
    $\begin{pmatrix}
        1 & 0 & 0 \\
        0 & 3 & 0 \\
        0 & 0 & 1 \\
    \end{pmatrix}$,
    $\begin{pmatrix}
        0 & 1/2 & 0
    \end{pmatrix}$ & Incompatible MPG \\ \\
    6 & \parbox{1.8cm}{$Cmc'm$ \\ (\#63.460)} & 
    $\begin{pmatrix}
        1 & 0 & 0 \\
        0 & 3 & 0 \\
        0 & 0 & 1 \\
    \end{pmatrix}$,
    $\begin{pmatrix}
        0 & 1/2 & 0
    \end{pmatrix}$ & Incompatible MPG \\ \\
    7 & \parbox{1.8cm}{$Cm'cm$ \\ (\#63.459)} & 
    $\begin{pmatrix}
        1 & 0 & 0 \\
        0 & 3 & 0 \\
        0 & 0 & 1 \\
    \end{pmatrix}$,
    $\begin{pmatrix}
        0 & 1/2 & 0
    \end{pmatrix}$ & Incompatible MPG \\ \\
    8 & \parbox{1.8cm}{$Cmcm$ \\ (\#63.457)} & 
    $\begin{pmatrix}
        1 & 0 & 0 \\
        0 & 3 & 0 \\
        0 & 0 & 1 \\
    \end{pmatrix}$,
    $\begin{pmatrix}
        0 & 1/2 & 0
    \end{pmatrix}$ & Incompatible MPG \\ \\
    9 & \parbox{1.8cm}{$Am'a'2$ \\ (\#40.207)} & 
    $\begin{pmatrix}
        0 & 0 & -1 \\
        0 & 3 & 0 \\
        1 & 0 & 0 \\
    \end{pmatrix}$,
    $\begin{pmatrix}
        0 & 1/2 & 0
    \end{pmatrix}$ & One finite $\rho_\mathrm{H}^0$ component \\ \\
    10 & \parbox{1.8cm}{$Ama'2'$ \\ (\#40.206)} & 
    $\begin{pmatrix}
        0 & 0 & -1 \\
        0 & 3 & 0 \\
        1 & 0 & 0 \\
    \end{pmatrix}$,
    $\begin{pmatrix}
        0 & 1/2 & 0
    \end{pmatrix}$ & One finite $\rho_\mathrm{H}^0$ component \\ \\
    11 & \parbox{1.8cm}{$Am'a2'$ \\ (\#40.205)} & 
    $\begin{pmatrix}
        0 & 0 & -1 \\
        0 & 3 & 0 \\
        1 & 0 & 0 \\
    \end{pmatrix}$,
    $\begin{pmatrix}
        0 & 1/2 & 0
    \end{pmatrix}$ & One finite $\rho_\mathrm{H}^0$ component \\ \\
    12 & \parbox{1.8cm}{$Ama2$ \\ (\#40.203)} & 
    $\begin{pmatrix}
        0 & 0 & -1 \\
        0 & 3 & 0 \\
        1 & 0 & 0 \\
    \end{pmatrix}$,
    $\begin{pmatrix}
        0 & 1/2 & 0
    \end{pmatrix}$ & Incompatible MPG \\ \\
    13 & \parbox{1.8cm}{$Am'm'2$ \\ (\#38.191)} & 
    $\begin{pmatrix}
        0 & 1 & 0 \\
        0 & 0 & 3 \\
        1 & 0 & 0 \\
    \end{pmatrix}$,
    $\begin{pmatrix}
        0 & 0 & 1/4
    \end{pmatrix}$ & One finite $\rho_\mathrm{H}^0$ component \\ \\
    14 & \parbox{1.8cm}{$Amm'2'$ \\ (\#38.190)} & 
    $\begin{pmatrix}
        0 & -1 & 0 \\
        0 & 0 & -3 \\
        1 & 0 & 0 \\
    \end{pmatrix}$,
    $\begin{pmatrix}
        0 & 0 & 1/4
    \end{pmatrix}$ & One finite $\rho_\mathrm{H}^0$ component \\ \\
    15 & \parbox{1.8cm}{$Am'm2'$ \\ (\#38.189)} & 
    $\begin{pmatrix}
        0 & 1 & 0 \\
        0 & 0 & 3 \\
        1 & 0 & 0 \\
    \end{pmatrix}$,
    $\begin{pmatrix}
        0 & 0 & 1/4
    \end{pmatrix}$ & \parbox{6cm}{\centering Incompatible magnetic \\ moment components} \\ \\
    16 & \parbox{1.8cm}{$Amm2$ \\ (\#38.187)} & 
    $\begin{pmatrix}
        0 & 1 & 0 \\
        0 & 0 & 3 \\
        1 & 0 & 0 \\
    \end{pmatrix}$,
    $\begin{pmatrix}
        0 & 0 & 1/4
    \end{pmatrix}$ & Incompatible MPG \\ \\
    17 & \parbox{1.8cm}{$Cm'c'2_1$ \\ (\#36.176)} & 
    $\begin{pmatrix}
        1 & 0 & 0 \\
        0 & 3 & 0 \\
        0 & 0 & 1 \\
    \end{pmatrix}$,
    $\begin{pmatrix}
        0 & 1/2 & 0
    \end{pmatrix}$ & One finite $\rho_\mathrm{H}^0$ component \\ \\
    18 & \parbox{1.8cm}{$Cmc'2_1'$ \\ (\#36.175)} & 
    $\begin{pmatrix}
        1 & 0 & 0 \\
        0 & 3 & 0 \\
        0 & 0 & 1 \\
    \end{pmatrix}$,
    $\begin{pmatrix}
        0 & 1/2 & 0
    \end{pmatrix}$ & \parbox{6cm}{\centering Incompatible magnetic \\ moment components} \\ \\
    19 & \parbox{1.8cm}{$Cm'c2_1'$ \\ (\#36.174)} & 
    $\begin{pmatrix}
        1 & 0 & 0 \\
        0 & 3 & 0 \\
        0 & 0 & 1 \\
    \end{pmatrix}$,
    $\begin{pmatrix}
        0 & 1/2 & 0
    \end{pmatrix}$ & One finite $\rho_\mathrm{H}^0$ component \\ \\
    20 & \parbox{1.8cm}{$Cmc2_1$ \\ (\#36.172)} & 
    $\begin{pmatrix}
        1 & 0 & 0 \\
        0 & 3 & 0 \\
        0 & 0 & 1 \\
    \end{pmatrix}$,
    $\begin{pmatrix}
        0 & 1/2 & 0
    \end{pmatrix}$ & Incompatible MPG \\ \\
    21 & \parbox{1.8cm}{$C22'2_1'$ \\ (\#20.34)} & 
    $\begin{pmatrix}
        0 & 1 & 0 \\
        -3 & 0 & 0 \\
        0 & 0 & 1 \\
    \end{pmatrix}$,
    $\begin{pmatrix}
        0 & 1/2 & 1/4
    \end{pmatrix}$ & One finite $\rho_\mathrm{H}^0$ component \\ \\
    22 & \parbox{1.8cm}{$C22'2_1'$ \\ (\#20.34)} & 
    $\begin{pmatrix}
        1 & 0 & 0 \\
        0 & 3 & 0 \\
        0 & 0 & 1 \\
    \end{pmatrix}$,
    $\begin{pmatrix}
        0 & 1/2 & 0
    \end{pmatrix}$ & One finite $\rho_\mathrm{H}^0$ component \\ \\
    23 & \parbox{1.8cm}{$C2'2'2_1$ \\ (\#20.33)} & 
    $\begin{pmatrix}
        1 & 0 & 0 \\
        0 & 3 & 0 \\
        0 & 0 & 1 \\
    \end{pmatrix}$,
    $\begin{pmatrix}
        0 & 1/2 & 0
    \end{pmatrix}$ & One finite $\rho_\mathrm{H}^0$ component \\ \\
    24 & \parbox{1.8cm}{$C222_1$ \\ (\#20.31)} & 
    $\begin{pmatrix}
        1 & 0 & 0 \\
        0 & 3 & 0 \\
        0 & 0 & 1 \\
    \end{pmatrix}$,
    $\begin{pmatrix}
        0 & 1/2 & 0
    \end{pmatrix}$ & Incompatible MPG \\ \\
    25 & \parbox{1.8cm}{$C2'/c'$ \\ (\#15.89)} & 
    $\begin{pmatrix}
        1 & 0 & 0 \\
        0 & 3 & 0 \\
        0 & 0 & 1 \\
    \end{pmatrix}$,
    $\begin{pmatrix}
        0 & 1/2 & 0
    \end{pmatrix}$ & Incompatible MPG \\ \\
    26 & \parbox{1.8cm}{$C2/c'$ \\ (\#15.88)} & 
    $\begin{pmatrix}
        1 & 0 & 0 \\
        0 & 3 & 0 \\
        0 & 0 & 1 \\
    \end{pmatrix}$,
    $\begin{pmatrix}
        0 & 1/2 & 0
    \end{pmatrix}$ & Incompatible MPG \\ \\
    27 & \parbox{1.8cm}{$C2'/c$ \\ (\#15.87)} & 
    $\begin{pmatrix}
        1 & 0 & 0 \\
        0 & 3 & 0 \\
        0 & 0 & 1 \\
    \end{pmatrix}$,
    $\begin{pmatrix}
        0 & 1/2 & 0
    \end{pmatrix}$ & Incompatible MPG \\ \\
    28 & \parbox{1.8cm}{$C2/c$ \\ (\#15.85)} & 
    $\begin{pmatrix}
        1 & 0 & 0 \\
        0 & 3 & 0 \\
        0 & 0 & 1 \\
    \end{pmatrix}$,
    $\begin{pmatrix}
        0 & 1/2 & 0
    \end{pmatrix}$ & Incompatible MPG \\ \\
    29 & \parbox{1.8cm}{$C2'/m'$ \\ (\#12.62)} & 
    $\begin{pmatrix}
        0 & -1 & 0 \\
        3 & 0 & 0 \\
        0 & 0 & 1 \\
    \end{pmatrix}$,
    $\begin{pmatrix}
        0 & 1/2 & 0
    \end{pmatrix}$ & Incompatible MPG \\ \\
    30 & \parbox{1.8cm}{$C2/m'$ \\ (\#12.61)} & 
    $\begin{pmatrix}
        0 & -1 & 0 \\
        3 & 0 & 0 \\
        0 & 0 & 1 \\
    \end{pmatrix}$,
    $\begin{pmatrix}
        0 & 1/2 & 0
    \end{pmatrix}$ & Incompatible MPG \\ \\
    31 & \parbox{1.8cm}{$C2'/m$ \\ (\#12.60)} & 
    $\begin{pmatrix}
        0 & -1 & 0 \\
        3 & 0 & 0 \\
        0 & 0 & 1 \\
    \end{pmatrix}$,
    $\begin{pmatrix}
        0 & 1/2 & 0
    \end{pmatrix}$ & Incompatible MPG \\ \\
    32 & \parbox{1.8cm}{$C2/m$ \\ (\#12.58)} & 
    $\begin{pmatrix}
        0 & -1 & 0 \\
        3 & 0 & 0 \\
        0 & 0 & 1 \\
    \end{pmatrix}$,
    $\begin{pmatrix}
        0 & 1/2 & 0
    \end{pmatrix}$ & Incompatible MPG \\ \\
    33 & \parbox{1.8cm}{$P2_1'/m'$ \\ (\#11.54)} & 
    $\begin{pmatrix}
        1 & 0 & -1/2 \\
        0 & 0 & -3/2 \\
        0 & 1 & 0 \\
    \end{pmatrix}$,
    $\begin{pmatrix}
        1/4 & -1/4 & 0
    \end{pmatrix}$ & Incompatible MPG \\ \\
    34 & \parbox{1.8cm}{$P2_1/m'$ \\ (\#11.53)} & 
    $\begin{pmatrix}
        1 & 0 & -1/2 \\
        0 & 0 & -3/2 \\
        0 & 1 & 0 \\
    \end{pmatrix}$,
    $\begin{pmatrix}
        1/4 & -1/4 & 0
    \end{pmatrix}$ & Incompatible MPG \\ \\
    35 & \parbox{1.8cm}{$P2_1'/m$ \\ (\#11.52)} & 
    $\begin{pmatrix}
        1 & 0 & -1/2 \\
        0 & 0 & -3/2 \\
        0 & 1 & 0 \\
    \end{pmatrix}$,
    $\begin{pmatrix}
        1/4 & -1/4 & 0
    \end{pmatrix}$ & Incompatible MPG \\ \\
    36 & \parbox{1.8cm}{$P2_1/m$ \\ (\#11.50)} & 
    $\begin{pmatrix}
        1 & 0 & -1/2 \\
        0 & 0 & -3/2 \\
        0 & 1 & 0 \\
    \end{pmatrix}$,
    $\begin{pmatrix}
        1/4 & -1/4 & 0
    \end{pmatrix}$ & Incompatible MPG \\ \\
    37 & \parbox{1.8cm}{$Cc'$ \\ (\#9.39)} & 
    $\begin{pmatrix}
        1 & 0 & 0 \\
        0 & 3 & 0 \\
        0 & 0 & 1 \\
    \end{pmatrix}$,
    $\begin{pmatrix}
        0 & 1/2 & 0
    \end{pmatrix}$ & Incorrect finite $\rho_\mathrm{H}^0$ components \\ \\
    38 & \parbox{1.8cm}{$Cc$ \\ (\#9.37)} & 
    $\begin{pmatrix}
        1 & 0 & 0 \\
        0 & 3 & 0 \\
        0 & 0 & 1 \\
    \end{pmatrix}$,
    $\begin{pmatrix}
        0 & 1/2 & 0
    \end{pmatrix}$ & One finite $\rho_\mathrm{H}^0$ component \\ \\
    39 & \parbox{1.8cm}{$Cm'$ \\ (\#8.34)} & 
    $\begin{pmatrix}
        0 & -1 & 0 \\
        3 & 0 & 0 \\
        0 & 0 & 1 \\
    \end{pmatrix}$,
    $\begin{pmatrix}
        0 & 0 & 0
    \end{pmatrix}$ & Compatible \\ \\
    40 & \parbox{1.8cm}{$Cm$ \\ (\#8.32)} & 
    $\begin{pmatrix}
        0 & -1 & 0 \\
        3 & 0 & 0 \\
        0 & 0 & 1 \\
    \end{pmatrix}$,
    $\begin{pmatrix}
        0 & 0 & 0
    \end{pmatrix}$ & \parbox{6cm}{\centering Incompatible magnetic \\ moment components} \\ \\
    41 & \parbox{1.8cm}{$Pm'$ \\ (\#6.20)} & 
    $\begin{pmatrix}
        1 & 0 & -1/2 \\
        0 & 0 & -3/2 \\
        0 & 1 & 0 \\
    \end{pmatrix}$,
    $\begin{pmatrix}
        0 & 0 & 1/4
    \end{pmatrix}$ & Incorrect finite $\rho_\mathrm{H}^0$ components \\ \\
    42 & \parbox{1.8cm}{$Pm$ \\ (\#6.18)} & 
    $\begin{pmatrix}
        1 & 0 & -1/2 \\
        0 & 0 & -3/2 \\
        0 & 1 & 0 \\
    \end{pmatrix}$,
    $\begin{pmatrix}
        0 & 0 & 1/4
    \end{pmatrix}$ & One finite $\rho_\mathrm{H}^0$ component \\ \\
    43 & \parbox{1.8cm}{$C2'$ \\ (\#5.15)} & 
    $\begin{pmatrix}
        1 & 0 & 0 \\
        0 & 3 & 0 \\
        0 & 0 & 1 \\
    \end{pmatrix}$,
    $\begin{pmatrix}
        0 & 0 & 1/4
    \end{pmatrix}$ & Incorrect finite $\rho_\mathrm{H}^0$ components \\ \\
    44 & \parbox{1.8cm}{$C2'$ \\ (\#5.15)} & 
    $\begin{pmatrix}
        0 & 1 & 0 \\
        -3 & 0 & 0 \\
        0 & 0 & 1 \\
    \end{pmatrix}$,
    $\begin{pmatrix}
        0 & 1/2 & 0
    \end{pmatrix}$ & In-plane toroidal moment \\ \\
    45 & \parbox{1.8cm}{$C2$ \\ (\#5.13)} & 
    $\begin{pmatrix}
        1 & 0 & 0 \\
        0 & 3 & 0 \\
        0 & 0 & 1 \\
    \end{pmatrix}$,
    $\begin{pmatrix}
        0 & 0 & 1/4
    \end{pmatrix}$ & One finite $\rho_\mathrm{H}^0$ component \\ \\
    46 & \parbox{1.8cm}{$C2$ \\ (\#5.13)} & 
    $\begin{pmatrix}
        0 & 1 & 0 \\
        -3 & 0 & 0 \\
        0 & 0 & 1 \\
    \end{pmatrix}$,
    $\begin{pmatrix}
        0 & 1/2 & 0
    \end{pmatrix}$ & One finite $\rho_\mathrm{H}^0$ component \\ \\
    47 & \parbox{1.8cm}{$P2_1'$ \\ (\#4.9)} & 
    $\begin{pmatrix}
        1 & 0 & -1/2 \\
        0 & 0 & -3/2 \\
        0 & 1 & 0 \\
    \end{pmatrix}$,
    $\begin{pmatrix}
        1/4 & -1/4 & 0
    \end{pmatrix}$ & Incorrect finite $\rho_\mathrm{H}^0$ components \\ \\
    48 & \parbox{1.8cm}{$P2_1$ \\ (\#4.7)} & 
    $\begin{pmatrix}
        1 & 0 & -1/2 \\
        0 & 0 & -3/2 \\
        0 & 1 & 0 \\
    \end{pmatrix}$,
    $\begin{pmatrix}
        1/4 & -1/4 & 0
    \end{pmatrix}$ & One finite $\rho_\mathrm{H}^0$ component \\ \\
    49 & \parbox{1.8cm}{$P\bar{1}'$ \\ (\#2.6)} & 
    $\begin{pmatrix}
        1 & -1/2 & 0 \\
        0 & 3/2 & 0 \\
        0 & 0 & 1 \\
    \end{pmatrix}$,
    $\begin{pmatrix}
        1/4 & -1/4 & 0
    \end{pmatrix}$ & Incompatible MPG \\ \\
    50 & \parbox{1.8cm}{$P\bar{1}$ \\ (\#2.4)} & 
    $\begin{pmatrix}
        1 & -1/2 & 0 \\
        0 & 3/2 & 0 \\
        0 & 0 & 1 \\
    \end{pmatrix}$,
    $\begin{pmatrix}
        1/4 & -1/4 & 0
    \end{pmatrix}$ & Incompatible MPG \\ \\
    51 & \parbox{1.8cm}{$P1$ \\ (\#1.1)} & 
    $\begin{pmatrix}
        1 & -1/2 & 0 \\
        0 & 3/2 & 0 \\
        0 & 0 & 1 \\
    \end{pmatrix}$,
    $\begin{pmatrix}
        0 & 0 & 0
    \end{pmatrix}$ & \parbox{6cm}{\centering Compatible, lower \\ symmetry than necessary} \\ \\
    \label{tab:cmcm}
\end{longtable}

\begin{longtable}[t]{llcc}
    \setlength{\tabcolsep}{3pt}\\
    \caption{Magnetic subgroups for $Pmm2$ with $\textbf{k}=(0,0,2/3)$.}
    \\ 
    \hline
    \\
    & MSG & Transformation & (In)compatibility \\
    & & \textbf{P}, \textbf{p} & \\ \\
    \hline
    \\
    1 & \parbox{1.8cm}{$Pm'm'2$ \\ (\#25.60)} & 
    $\begin{pmatrix}
        1 & 0 & 0 \\
        0 & 1 & 0 \\
        0 & 0 & 3 \\
    \end{pmatrix}$,
    $\begin{pmatrix}
        0 & 0 & 0
    \end{pmatrix}$ & One finite $\rho_\mathrm{H}^0$ component \\ \\
    2 & \parbox{1.8cm}{$Pm'm2'$ \\ (\#25.59)} & 
    $\begin{pmatrix}
        1 & 0 & 0 \\
        0 & 1 & 0 \\
        0 & 0 & 3 \\
    \end{pmatrix}$,
    $\begin{pmatrix}
        0 & 0 & 0
    \end{pmatrix}$ & One finite $\rho_\mathrm{H}^0$ component \\ \\
    3 & \parbox{1.8cm}{$Pm'm2'$ \\ (\#25.59)} & 
    $\begin{pmatrix}
        0 & 1 & 0 \\
        -1 & 0 & 0 \\
        0 & 0 & 3 \\
    \end{pmatrix}$,
    $\begin{pmatrix}
        0 & 0 & 0
    \end{pmatrix}$ & \parbox{6cm}{\centering Incompatible magnetic \\ moment components} \\ \\
    4 & \parbox{1.8cm}{$Pmm2$ \\ (\#25.57)} & 
    $\begin{pmatrix}
        1 & 0 & 0 \\
        0 & 1 & 0 \\
        0 & 0 & 3 \\
    \end{pmatrix}$,
    $\begin{pmatrix}
        0 & 0 & 0
    \end{pmatrix}$ & Incompatible MPG \\ \\
    5 & \parbox{1.8cm}{$Pm'$ \\ (\#6.20)} & 
    $\begin{pmatrix}
        1 & 0 & 0 \\
        0 & 1 & 0 \\
        0 & 0 & 3 \\
    \end{pmatrix}$,
    $\begin{pmatrix}
        0 & 0 & 0
    \end{pmatrix}$ & Incorrect finite $\rho_\mathrm{H}^0$ components\\ \\
    6 & \parbox{1.8cm}{$Pm'$ \\ (\#6.20)} & 
    $\begin{pmatrix}
        0 & 1 & 0 \\
        -1 & 0 & 0 \\
        0 & 0 & 3 \\
    \end{pmatrix}$,
    $\begin{pmatrix}
        0 & 0 & 0
    \end{pmatrix}$ & Compatible \\ \\
    7 & \parbox{1.8cm}{$Pm$ \\ (\#6.18)} & 
    $\begin{pmatrix}
        1 & 0 & 0 \\
        0 & 1 & 0 \\
        0 & 0 & 3 \\
    \end{pmatrix}$,
    $\begin{pmatrix}
        0 & 0 & 0
    \end{pmatrix}$ & One finite $\rho_\mathrm{H}^0$ component \\ \\
    8 & \parbox{1.8cm}{$Pm$ \\ (\#6.18)} & 
    $\begin{pmatrix}
        0 & 1 & 0 \\
        -1 & 0 & 0 \\
        0 & 0 & 3 \\
    \end{pmatrix}$,
    $\begin{pmatrix}
        0 & 0 & 0
    \end{pmatrix}$ & \parbox{6cm}{\centering Incompatible magnetic \\ moment components} \\ \\
    9 & \parbox{1.8cm}{$P2'$ \\ (\#3.3)} & 
    $\begin{pmatrix}
        1 & 0 & 0 \\
        0 & 0 & 1 \\
        0 & -3 & 0 \\
    \end{pmatrix}$,
    $\begin{pmatrix}
        0 & 0 & 0
    \end{pmatrix}$ & Incorrect finite $\rho_\mathrm{H}^0$ components \\ \\
    10 & \parbox{1.8cm}{$P2$ \\ (\#3.1)} & 
    $\begin{pmatrix}
        1 & 0 & 0 \\
        0 & 0 & 1 \\
        0 & -3 & 0 \\
    \end{pmatrix}$,
    $\begin{pmatrix}
        0 & 0 & 0
    \end{pmatrix}$ & One finite $\rho_\mathrm{H}^0$ component \\ \\
    11 & \parbox{1.8cm}{$P1$ \\ (\#1.1)} & 
    $\begin{pmatrix}
        1 & 0 & 0 \\
        0 & 1 & 0 \\
        0 & 0 & 3 \\
    \end{pmatrix}$,
    $\begin{pmatrix}
        0 & 0 & 0
    \end{pmatrix}$ & \parbox{6cm}{\centering Compatible, lower \\ symmetry than necessary} \\ \\
    \label{tab:pmm2}
\end{longtable}

\begin{figure}[h]
    \centering
    \subfloat{\includegraphics[width=0.47\columnwidth]{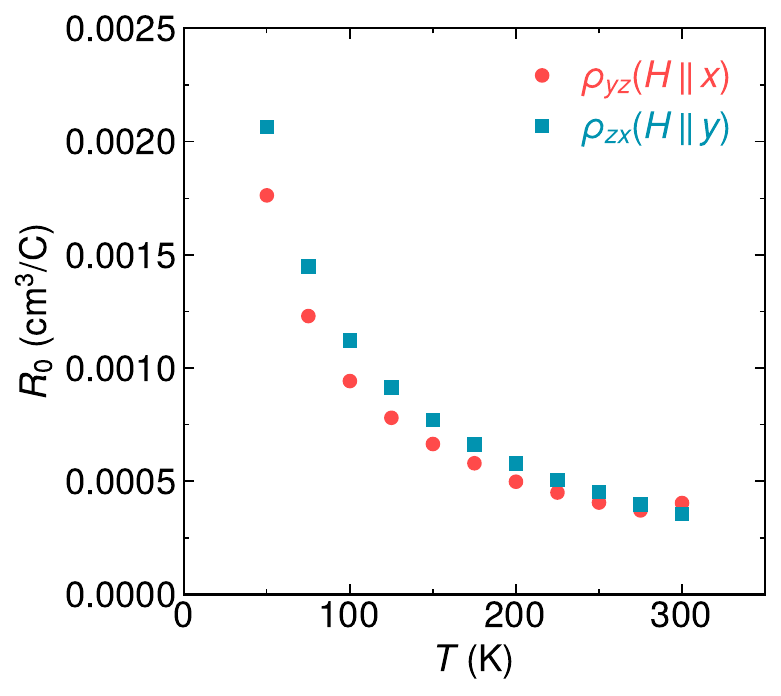}}
    \hspace{2em}
    \subfloat{\includegraphics[width=0.45\columnwidth]{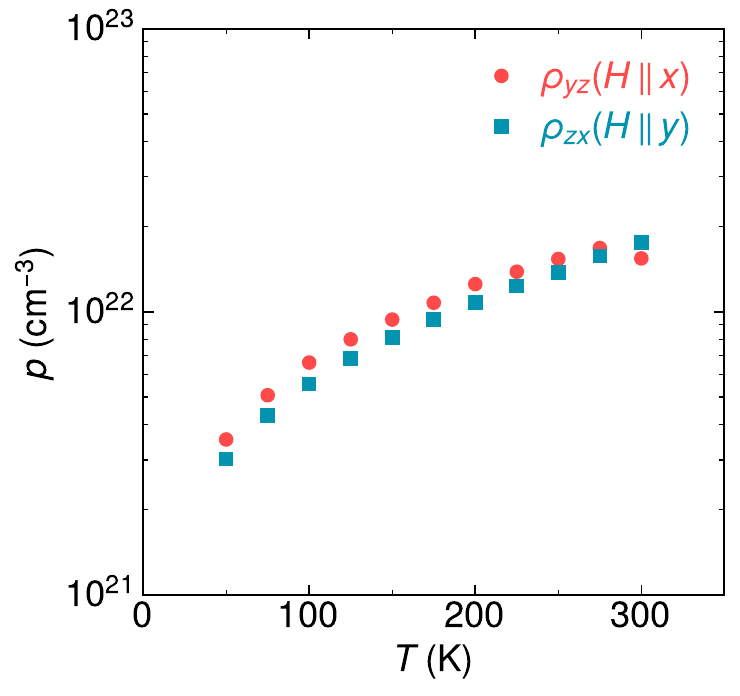}}
    \caption{Paramagnetic Hall resistivity data are fit to a single carrier model to give an ordinary Hall coefficient ($R_\mathrm{0}$) from which a carrier concentration ($p$) is extracted.}
    \label{fig:single_band}
\end{figure}

Figure~\ref{fig:single_band} shows single-band fits [$\rho_\mathrm{H}=R_\mathrm{0}\mu_\mathrm{0}H$] for temperatures where the Hall resistivity is linear in the paramagnetic regime. $R_\mathrm{0}$ is converted to a hole carrier concentration ($p$) by the relationship $p=1/(R_\mathrm{0}q)$, where $q$ is the elementary charge.

